\documentclass[12pt]{article}
\pdfoutput=1
\usepackage{jheppub}

\usepackage{multirow}
\usepackage{amsmath,amssymb,graphicx,verbatim} 
\usepackage{hyperref}
\usepackage{enumerate}
\usepackage{array}
\usepackage{caption}
\usepackage{subcaption}
\usepackage{units}

\newcommand{\figref}[1]{Fig.~\ref{fig:#1}}
\newcommand{\figrefs}[2]{Figs.\ \ref{fig:#1}, \ref{fig:#2}}
\renewcommand{\eqref}[1]{Eq.~(\ref{eq:#1})}

\newcommand{\aref}[1]{Appendix \ref{a.#1}}
\newcommand{\sref}[1]{Section \ref{s.#1}}
\newcommand{\ssref}[1]{Section \ref{ss.#1}}

\newcommand{\tref}[1]{Table \ref{t.#1}}
\newcommand{\tsref}[2]{Tabs.~\ref{t.#1} and \ref{t.#2}}

\def\gev{\, {\rm GeV}}
\def\tev{\, {\rm TeV}}

\newcommand\MadGraph{{\sc\small MadGraph}}
\newcommand\MadGraphFull{{\sc\small MadGraph5\_aMC@NLO}}
\newcommand\Pythia{{\sc\small Pythia~8}}
\newcommand\PythiaSix{{\sc\small Pythia~6}}
\newcommand\Delphes{{\sc\small Delphes~3}}
\newcommand\DelphesFull{{\sc\small Delphes~v3.4.1}}
\newcommand\MadSpin{{\sc\small MadSpin}}
\newcommand\FastJet{{\sc\small FastJet}}

\newcommand\PowhegBox{{\sc\small Powheg-Box}}

\def\doubleq#1{``{#1}"}

\def\E#1{\times 10^{#1}}

\title{Measurement of the Triple Higgs Coupling at a HE-LHC}

\abstract{
The currently unmeasured triple Higgs coupling is one of the strong motivations for future physics programs at the LHC and beyond.  A sufficiently precise measurement can lead to qualitative changes in our understanding of electroweak symmetry breaking and the cosmological history of the Higgs potential.  As such, the quantitative measurement of this coupling is now one of the benchmark measurements for any proposed collider.  We study the capability of a potential $27\tev$ HE-LHC upgrade in measuring the Higgs trilinear coupling via the di-Higgs production process in the $b\bar{b}\gamma\gamma$ channel. We emphasize that a key background from single Higgs production via gluon fusion has been underestimated and underappreciated in prior studies.  We perform a detailed study taking into account two different potential detector scenarios, and validate against HL-LHC projections from ATLAS.  We find that the di-Higgs production process can be observed at $\geq 4.5 \sigma$, corresponding to a $\sim 40\%$ measurement of the Higgs self-coupling, with $15\,\mathrm{ab}^{-1}$ of data at the HE-LHC.
}

\author[a,b]{Samuel Homiller}
\author[a]{and Patrick Meade}

\affiliation[a]{C. N. Yang Institute for Theoretical Physics\\ Stony Brook University, Stony Brook, NY 11794}
\affiliation[b]{Department of Physics,\\ Brookhaven National Laboratory, Upton, NY 11973}

\emailAdd{samuel.homiller@insti.physics.sunysb.edu}
\emailAdd{meade@insti.physics.sunysb.edu}

\begin{document}

\begin{flushright}
\small{YITP-SB-18-33}
\end{flushright}

\maketitle

\section{Introduction}
\label{s.intro} \setcounter{equation}{0} \setcounter{footnote}{0}

Since the discovery of the Higgs boson~\cite{Aad:2012tfa, Chatrchyan:2012xdj}, one of the primary goals in high energy physics has been to understand its properties better.  While the Higgs looks very similar to the Standard Model (SM) Higgs, many of its properties have yet to been measured, or measured sufficiently well to test for many possibilities of beyond the SM (BSM) physics.  One of the most exciting possibilities is reaching the level of precision where quantitative measurements can lead to qualitative changes in our understanding of electroweak symmetry breaking (EWSB). 

While the quantitative improvement in Higgs couplings to other particles has been studied in many contexts at future colliders, e.g. naturalness~\cite{Fan:2014txa, Essig:2017zwe}, the most exciting possibility is the measurement of the Higgs self interaction.  Such an interaction has never been observed in nature and it is the only direct window to further information about the Higgs potential itself.  Currently we only have experimentally determined the location of a minimum of the Higgs potential and the value of the second derivative at that minimum.  To learn more about the nature of the Higgs and its cosmological history requires the measurement of the triple Higgs coupling and beyond.  Unfortunately, measuring the Higgs self-coupling at the $14\tev$ LHC appears exceedingly difficult unless its value deviates substantially from the Standard Model prediction~\cite{ATL-PHYS-PUB-2014-019, Goertz:2014qta, Cao:2015oaa, ATL-PHYS-PUB-2017-001, Kim:2018uty}, although there is potential room for incremental improvement~\cite{Goertz:2013kp, Kim:2018cxf}.  Unfortunately, simply measuring the coupling is not enough to necessarily change our qualitative understanding of the Higgs potential or its cosmological evolution.  For example, the strength of a potential EW phase transition (EWPT) depends critically on the value of the effective triple Higgs coupling, and must be measured to the $\mathcal{O}(10\%)$ level or better to distinguish the order of the phase transition in some cases~\cite{Chung:2012vg, Curtin:2014jma}. Furthermore it has been recently shown that there might not even be an EW phase transition at typical EW scales, or even possibly at any scale~\cite{Meade:2018saz}. In these cases the triple Higgs coupling can distinguish the qualitatively different scenario, but an even more precise measurement is needed than to distinguish a first-order from second-order EWPT.  Therefore the actual quantitative precision a future collider program can reach in this coupling is of paramount importance.

With planning for the High-Luminosity phase of the LHC underway, there is growing interest in the possibility of an upgrade to the center of mass energy of the LHC. Such an upgrade, known as High-Energy (HE) LHC, would reach a center of mass energy of $27\tev$, and may be the next opportunity to explore physics at the energy frontier.
While the measurement of the self-coupling has been well studied at a potential 100 TeV collider~\cite{Yao:2013ika, Barr:2014sga, He:2015spf, Azatov:2015oxa, Contino:2016spe, Cao:2016zob, Kling:2016lay,Banerjee:2018yxy, Chang:2018uwu} 
and proposed lepton colliders~\cite{Asner:2013psa,Barklow:2017awn,McCullough:2013rea}, 
there are far fewer studies for the HE-LHC~\cite{Goncalves:2018yva}.  In particular, despite the extensive work for 100 TeV hadron colliders, the relevant backgrounds are still being understood and better estimated.  In this work we carefully analyze the HE-LHC measurement of the triple Higgs coupling from di-Higgs production in the $b\bar{b}\gamma\gamma$ channel. We implement this study using a \Delphes~\cite{DeFavereau:2013fsa} based detector simulation and two possible parametrizations for ECAL resolution.  We validate our study against existing ATLAS results as described in the Appendix and provide $14\tev$ projections alongside those for HE-LHC. 
Contrary to some existing studies for future hadron colliders, we find that the largest background, after typical analysis cuts to avoid the $t\bar{t}h$ background, comes from single Higgs production via gluon fusion in association with additional jets.  We study this background from a number of different perspectives and emphasize that this background increases in importance as one moves to higher energy colliders. 
In our study we do not optimize for observation of di-Higgs production alone, as could be done, as sensitivity to anomalous couplings is the primary focus for elucidating properties of the Higgs potential.

The rest of the paper is structured as follows, in Section 2, we discuss the details of di-Higgs production via gluon fusion at a hadron collider, focusing in particular on the decay channel $hh \to b\bar{b}\gamma\gamma$ and the simulation of the signal and backgrounds. Section 3 contains more information on the HE-LHC scenario and the detector simulation used for our study. In Section 4 we describe our cut-and-count analysis and the significance and precision attainable on the di-Higgs production cross section and trilinear coupling, as well as several benchmark figures on the capabilities of an HE-LHC compared to the High Luminosity (HL-LHC) scenario currently planned. We summarize our results in Section 5. In the Appendix we demonstrate the validation of our methods through a comparison with the ATLAS projection for this channel at the HL-LHC~\cite{ATL-PHYS-PUB-2017-001}.

\section{Di-Higgs Signal and Backgrounds}
\label{s.methods} \setcounter{equation}{0} \setcounter{footnote}{0}

\subsection{Di-Higgs Production}

\begin{figure}[ht]
\centering
\begin{subfigure}{0.4\textwidth}
  \centering
  \includegraphics[width=.9\linewidth]{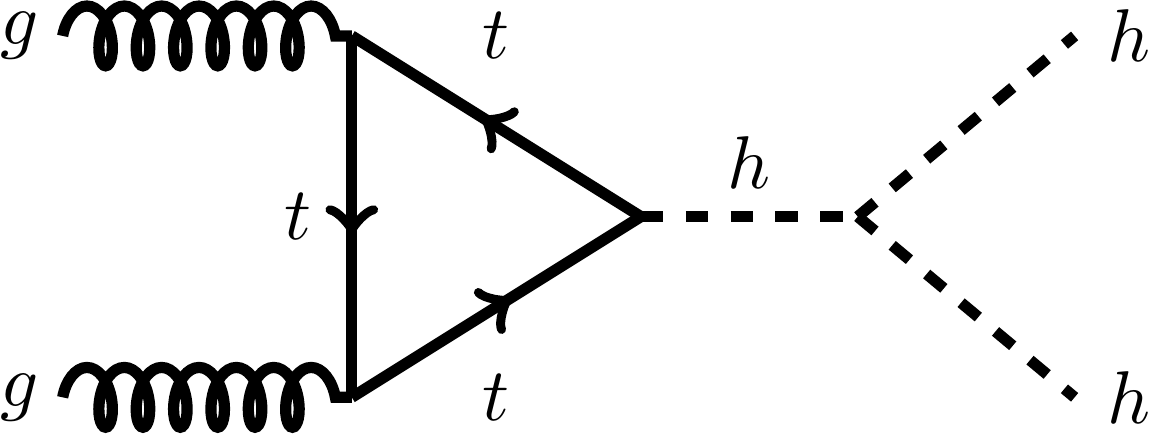}
\end{subfigure}
\hspace{1cm}
\begin{subfigure}{0.4\textwidth}
  \centering
  \includegraphics[width=.9\linewidth]{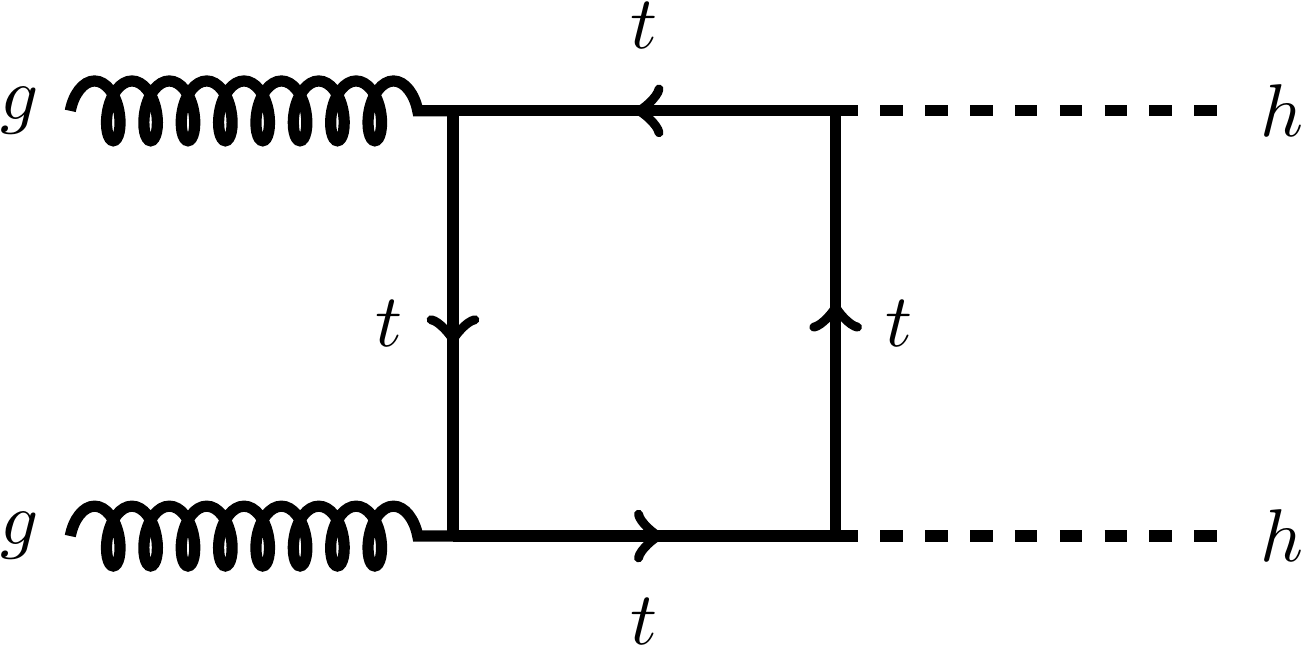}
\end{subfigure}
\caption{Lowest order Feynman diagrams contributing to di-Higgs production via gluon-gluon fusion. An additional box diagram with crossed final states is not shown.}\label{fig:feynmandiags}
\end{figure}

The most direct way to measure the Higgs trilinear coupling, $\lambda_3$, at a hadron collider is via the Higgs pair production process, which arises primarily from $gg \rightarrow hh$.  Our conventions are chosen such that after EWSB the interactions from the Higgs potential are given by
\begin{equation}\label{eq:v_int}
  V_{\text{int}} = \lambda_3\frac{m_h^2}{2v}h^3 + \lambda_4\frac{m_h^2}{8v^2}h^4
\end{equation}
so that $\lambda_3 = \lambda_4 = 1$ in the SM and $m_h = \sqrt{2\lambda v^2}$ is the physical Higgs boson mass.  The lowest order diagrams contributing to di-Higgs production for $gg \to hh$ in the SM, shown in \figref{feynmandiags}, arise from the \doubleq{triangle} diagram, as in single-Higgs production with an additional $hhh$ vertex, and from the \doubleq{box} diagram, which is independent of $\lambda_3$. To leading order, these amplitudes scale as~\cite{Contino:2016spe}:
\begin{equation}\centering
  \mathcal{M}_{\triangle} \sim \lambda_3 \frac{\alpha_s}{4\pi}y_t \frac{m_h^2}{\hat{s}} \left(\log \frac{m_t^2}{\hat{s}} + i\pi\right)^2,
  \qquad
  \mathcal{M}_{\square} \sim \frac{\alpha_s}{4\pi}y_t^2.
\label{eq:dihiggs_amplitudes}
\end{equation}
Due to the additional fermion line in the box diagram, the two amplitudes interfere destructively, leading to a $\sim 50\%$ reduction in the total cross section in the Standard Model, with a maximum cancellation near $\lambda_3 \sim 2$.\\

In this study, we focus exclusively on the gluon-gluon fusion channel of di-Higgs production, which dominates at all relevant energies~\cite{Frederix:2014hta}\footnote{See however, refs. \cite{Arganda:2018ftn, Cao:2015oxx}.}. Only the leading-order (LO) cross section was known exactly for many years~\cite{Eboli:1987dy, Glover:1987nx, Plehn:1996wb}, with a number of approximations to account for higher-order corrections.
These include next-to-leading order (NLO)~\cite{Dawson:1998py} and NNLO~\cite{deFlorian:2013jea, deFlorian:2016uhr} corrections in QCD, using the infinite top quark mass limit, as well as estimates of threshold resummation effects at NNLL accuracy~\cite{deFlorian:2015moa, Shao:2013bz}.
Recently, the cross section was computed exactly at NLO, including all top-quark mass effects~\cite{Borowka:2016ehy, Borowka:2016ypz} and matched to parton showers~\cite{Heinrich:2017kxx}. A better approximation can thus be obtained by shifting the NNLO+NNLL values in~\cite{DeFlorian:2016spz} to account for the finite-mass effects, leading to the values shown in \tref{dihiggs_csx} from~\cite{Grazzini:2018bsd}.

\begingroup
\renewcommand*{\arraystretch}{1.3}
\begin{table}[ht]
\centering
\begin{tabular}{|c|c|c|c|}
\hline
                                     						& $14\tev$ 	& $27\tev$ 	& $100\tev$ \\ \hline
$\sigma_{gg\rightarrow hh}$ [fb]     	& $39.58^{+1.4\%}_{-4.7\%}$  	& $154.2^{+0.7\%}_{-3.8\%}$  	& $1406^{+0.5\%}_{-2.8\%}$    \\ \hline
\end{tabular}
\caption{The Standard Model cross-section for di-Higgs production via gluon-gluon fusion at several different center of mass energies, based on the ``Born-projected approximation" in~\cite{Grazzini:2018bsd}. Scale uncertainties are shown as superscripts/subscripts.}
\label{t.dihiggs_csx}
\end{table}
\endgroup

\subsection{Signal and Primary Backgrounds}

There are a number of decay channels available for $hh$ production, many of which can be exploited by a future collider when the production rate is relatively large. 
These include $b\bar{b}\gamma\gamma$, $b\bar{b}b\bar{b}$, $b\bar{b}\tau^+\tau^-$, $b\bar{b}4\ell$, and $b\bar{b}\ell^+\ell^-$, where the leptons can arise either directly from the Higgs decay, or from an intermediate $W/Z$ (see \tref{dihiggs_brs} for a table of some common $hh$ final state branching ratios). Here, we focus on the $hh\rightarrow b\bar{b}\gamma\gamma$ decay channel ($BR = 0.264\%$ in the SM), which allows reasonable control of the backgrounds while maintaining a high enough rate from the large $h\rightarrow b\bar{b}$ branching ratio.
This channel has been shown to give the highest significance on the signal at future colliders~\cite{Contino:2016spe}, although the $hh\rightarrow b\bar{b}b\bar{b}$ prospects are competitive at $14\tev$~\cite{deLima:2014dta, ATL-PHYS-PUB-2016-024} 
and other final states may be interesting as well~\cite{Papaefstathiou:2012qe, Papaefstathiou:2015iba, Adhikary:2017jtu, Aaboud:2018sfw}.

\begingroup
\renewcommand*{\arraystretch}{1.2}
\begin{table}[ht]
\centering
\begin{tabular}{|c | c c c c c |}
\hline
				& $b\bar{b}$	& $\tau^+\tau^-$ 	& $WW (2\ell)$	& $\gamma\gamma$		& $ZZ(4\ell)$	\\ \hline
$b\bar{b}$			& $33.9\%$	& $7.31\%$		& $1.23\%$	& $0.264\%$			& $0.0144\%$\\
$\tau^+\tau^-$		&			& $0.393\%$		& $0.132\%$	& $0.0285\%$			& $1.6\E{-5}$	\\
$WW (2\ell)$		&			&				& $0.011\%$	& $4.8\E{-5}$			& $2.6\E{-6}$	\\
$\gamma\gamma$	&			&				&			& $5.2\E{-6}$			& $5.6\E{-7}$	\\
$ZZ(4\ell)$		&			&				&			&					& $1.5\E{-8}$	\\ \hline
\end{tabular}
\caption{Branching ratios for some important $hh$ decay modes (based on~\cite{DeFlorian:2016spz}), where the first (second) higgs decay is shown in the row (column). The entries are symmetric along the diagonal.}
\label{t.dihiggs_brs}
\end{table}
\endgroup

Backgrounds to the $hh\rightarrow b\bar{b}\gamma\gamma$ mode arise from the single higgs production modes, $ggF(\gamma\gamma)$ (where the $b\bar{b}$ pair is produced by extra radiation), $Zh(\gamma\gamma)$, and $t\bar{t}h(\gamma\gamma)$.
There are also a number of non-resonant backgrounds arising from processes such as $b\bar{b}\gamma\gamma$, $jj\gamma\gamma$, $b\bar{b}j\gamma$, $c\bar{c}j\gamma$, $b\bar{b}jj$, and $Z(b\bar{b})\gamma\gamma$, where $j$ refers to a jet arising from light ($u, d, s, c$) quarks or gluons.
Finally, there are also backgrounds due to top-pair production, e.g. $t\bar{t}$ and $t\bar{t}\gamma$, where the additional photons can arise from either misidentified light jets or electrons. We neglect backgrounds where more than two misidentified particles are required, such as $t\bar{t}\gamma\gamma$ and $bjjj$, as their contributions are expected to be negligible.

\subsection{Signal and Background Simulations}\label{ss.simulations}

\begingroup
\renewcommand*{\arraystretch}{1.1}
\begin{table}[ht]
\centering
\begin{tabular}{|c|c|cc|c|}
\hline
\multirow{2}{*}{Process}     & \multirow{2}{*}{Generator} & \multicolumn{2}{c|}{$\sigma \cdot BR$ {[}fb{]}} & \multirow{2}{*}{Order QCD} \\ \cline{3-4}
                              						&									& $14\tev$	& $27\tev$		& \\ \hline
$h(b\bar{b})h(\gamma\gamma)$	& \MadGraph/\Pythia 	& 0.11			& 0.41				& NLO$^{*}$ \\ \hline
$t\bar{t}h(\gamma\gamma)$			& \Pythia						& 1.40			& 6.54				& NLO \\
$Zh(\gamma\gamma)$					& \Pythia						& 2.24			& 5.58				& NLO \\
$ggF(\gamma\gamma)$				& \MadGraph/\Pythia 	& 83.2			& 335.1				& LO$^{\dagger}$\\ \hline
$b\bar{b}\gamma\gamma$			& \MadGraph/\Pythia 	& $3.4\E{2}$	& $9.5\E{2}$     & LO \\
$c\bar{c}\gamma\gamma$        		& \MadGraph/\Pythia 	& $4.4\E{2}$	& $1.5\E{3}$     & LO \\
$jj\gamma\gamma$						& \MadGraph/\Pythia 	& $5.9\E{3}$	& $1.4\E{4}$     & LO \\
$b\bar{b}j\gamma$						& \MadGraph/\Pythia 	& $1.1\E{6}$	& $3.4\E{6}$     & LO \\
$c\bar{c}j\gamma$						& \MadGraph/\Pythia 	& $4.8\E{5}$	& $1.6\E{6}$     & LO \\
$b\bar{b}jj$									& \MadGraph/\Pythia 	& $3.7\E{8}$	& $1.5\E{9}$     & LO \\
$Z(b\bar{b})\gamma\gamma$		& \MadGraph/\Pythia 	& 2.61			& 5.23				& LO \\ \hline
$t\bar{t}$										& \MadGraph/\Pythia 	& $6.7\E{5}$	& $2.9\E{6}$     	& NNLO \\
$t\bar{t}\gamma$							& \MadGraph/\Pythia 	& $1.7\E{3}$	& $7.9\E{3}$     & NLO \\ \hline
\end{tabular}
\caption{List of signal and background processes and the event generator used to simulate their matrix elements and parton showering. Also shown are the cross sections of each process and the corresponding order in QCD at which the cross section used to normalize the expected event rate was computed. Note that we use the same naming scheme for backgrounds as ATLAS, but we do not list $b\bar{b}h$ as it is included in $ggF$ as described in the text.\\
$^{*}$The $hh$ signal is scaled to NLO in QCD with the full $m_t$ dependence, but also includes NNLO corrections in the $m_t \to \infty$ limit as well as threshold resummation effects to NNLL.\\
$^{\dagger}$The $ggF$ background was produced at leading order and in the $m_t \to \infty$ limit, but with two extra real emissions included. See text for details.
}
\label{t.background_csx}
\end{table}
\endgroup

\begin{figure}[!htbp]
\centering
\begin{subfigure}{.49\textwidth}
	\centering
	\includegraphics[width=\linewidth]{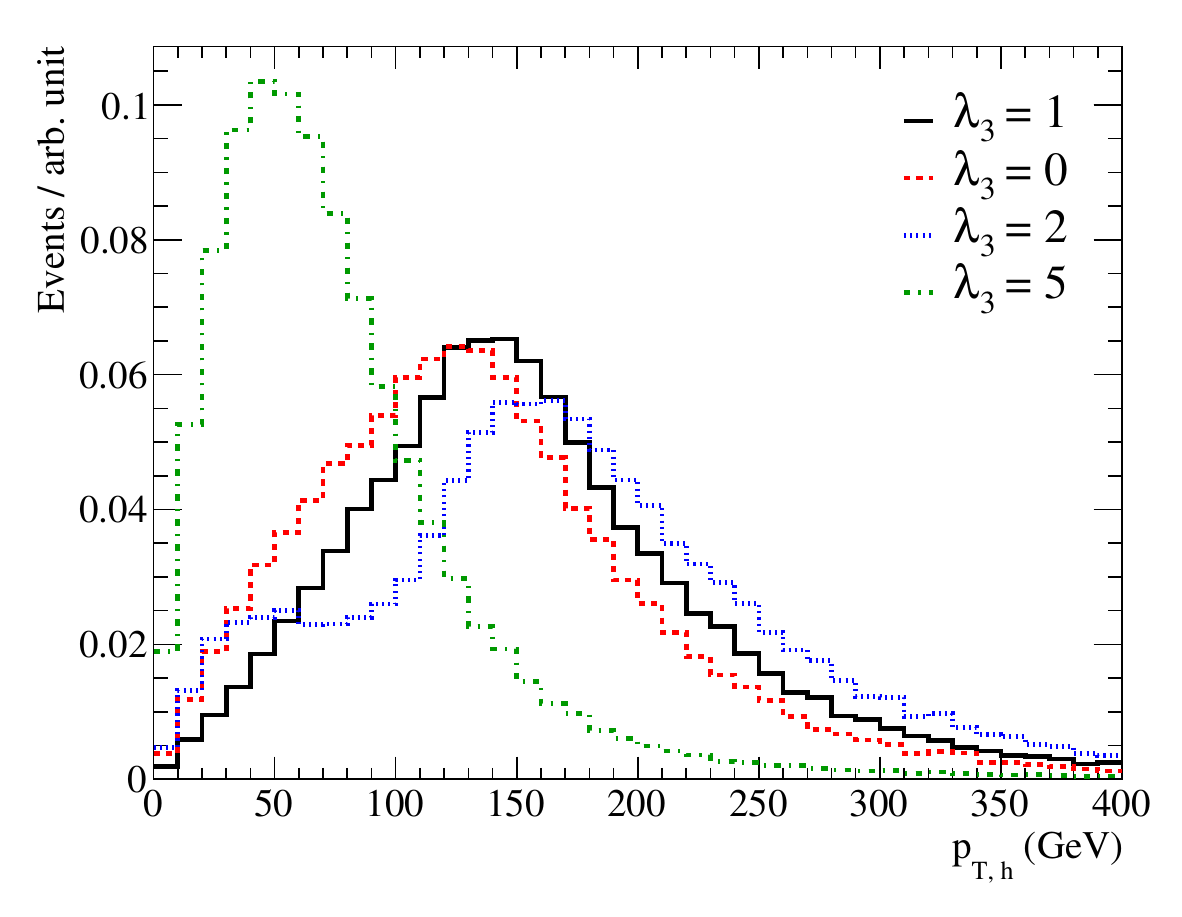}
	\caption{}\label{fig:dihiggs_kinematics1}
\end{subfigure}
\begin{subfigure}{.49\textwidth}
	\centering
	\includegraphics[width=\linewidth]{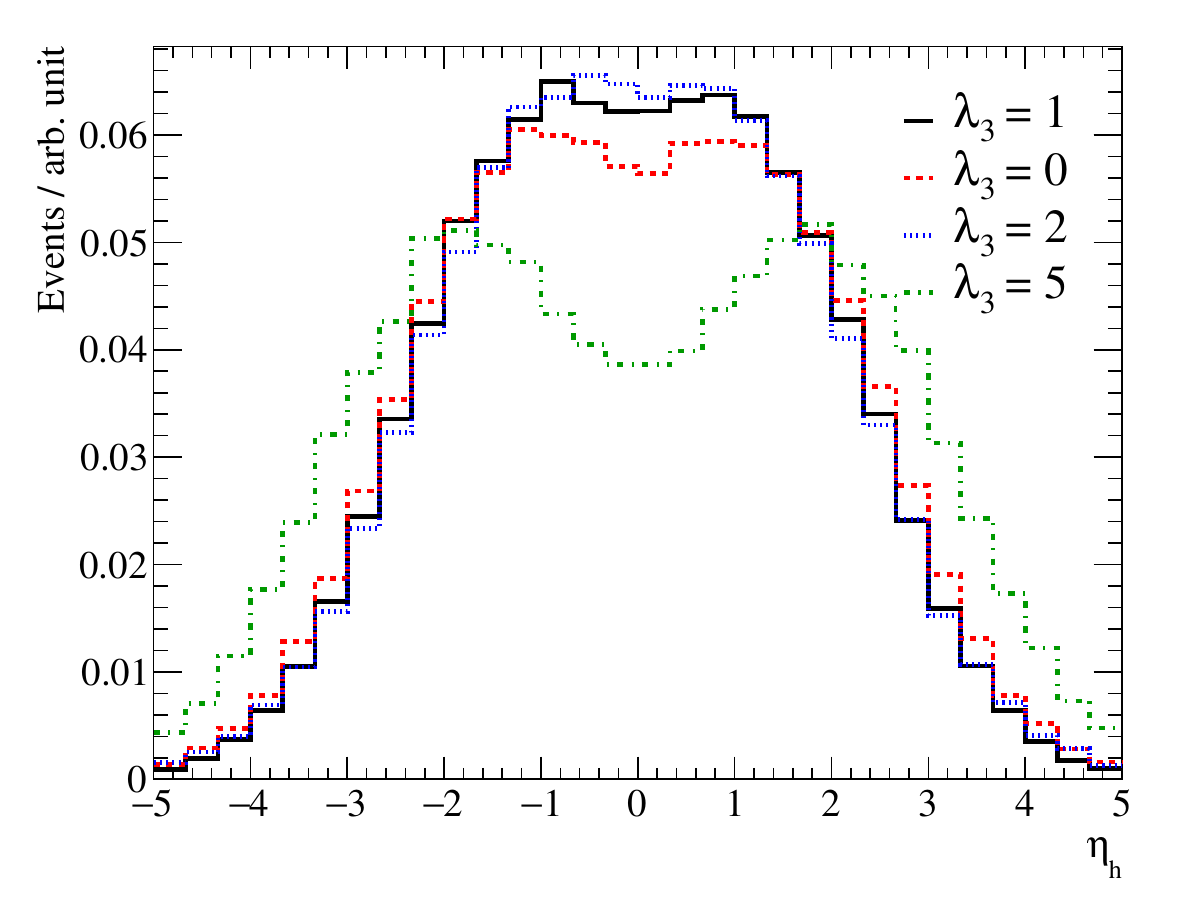}
	\caption{}\label{fig:dihiggs_kinematics2}
\end{subfigure}\\[1ex]
	\begin{subfigure}{\linewidth}
	\centering
	\includegraphics[width=0.5\linewidth]{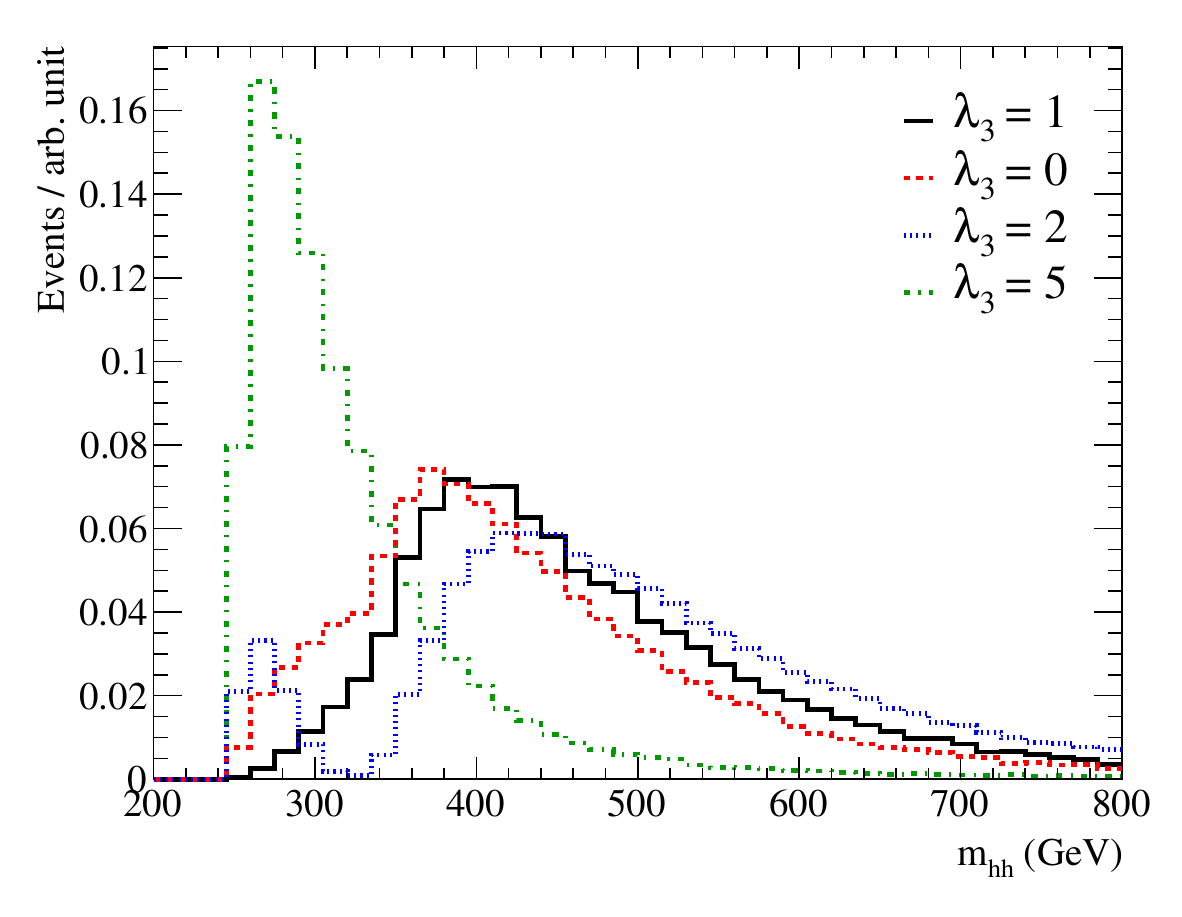}
	\caption{}\label{fig:dihiggs_kinematics3}
\end{subfigure}
\caption{(a): The transverse-momentum distribution of the true Higgs bosons generated in our $27\tev$ samples, prior to showering and detector smearing, for several different values of $\lambda_3$. (b): The same, but for the Higgs pseudorapidity. (c): The same, but for the distribution of the true Higgs pair invariant mass.}\label{fig:higgs_parton}
\end{figure}

For the signal simulation, the leading order loop level process is simulated directly using \MadGraphFull~\cite{Alwall:2014hca, Hirschi:2015iia} using the NNPDF2.3LO PDF set~\cite{Ball:2014uwa} including all finite top mass effects.
The \MadSpin\ package~\cite{Artoisenet:2012st} was used for the Higgs boson decays and \Pythia~\cite{Sjostrand:2014zea} for the showering and hadronization of events.
The LO signal is normalized to match the NLO+NNLO/NNLL cross sections listed in \tref{dihiggs_csx}. Signal events were also generated with the trilinear coupling $\lambda_3$ (defined as in \eqref{v_int}) modified to values ranging from $-1$ to $10$ times the SM value.
Some kinematics features of the di-Higgs signal at parton level are illustrated in \figref{higgs_parton} for different values of $\lambda_3$. Several of these features can be understood from the naive scaling in \eqref{dihiggs_amplitudes}; in particular, for large values of $\lambda_3$, the process becomes dominated by the \doubleq{triangle} diagram, which falls off for large $\hat{s}$ as clearly demonstrated in \figref{higgs_parton}.

The background from single-Higgs production via gluon fusion, $ggF(\gamma\gamma)$, was ignored in a number of previous projections, but, as demonstrated in \sref{results}, it is one of the dominant backgrounds to di-Higgs production. As such, we'll treat the generation of this background with special care in \ssref{ggf_background}. 
Events from other single Higgs production modes --- $Zh(\gamma\gamma)$ and $t\bar{t}h(\gamma\gamma)$\footnote{The associated $b\bar{b}h$ production mode usually considered is here included as part of the $ggF$ background. See \ssref{ggf_background} for details.} --- were generated directly in \Pythia\ at LO using the CTEQ6L1 PDF sets~\cite{Pumplin:2002vw}. 
To account for higher order effects, the $14\tev$ samples were normalized to the NLO cross sections based on the LHC Higgs Cross Section working group recommendations~\cite{DeFlorian:2016spz}. These recommendations were also used to compute $k$-factors at $7$, $8$, $13$ and $14\tev$. These $k$-factors were found to vary modestly as a function of energy, so we extrapolated to $27\tev$ based on a second-order polynomial in $\sqrt{E}$. 
The extrapolated $k$-factors are found to vary only by $4\%$ and $6\%$ for $t\bar{t}h$ and $Zh$ respectively.
Since these differences are relatively mild, we use these $k$-factors to normalize the LO samples, resulting in the cross sections presented in \tref{background_csx}.

Non-resonant background events were generated with \MadGraphFull~\cite{Alwall:2014hca} using the CTEQ6L1 PDF set and interfaced with \Pythia~\cite{Sjostrand:2014zea} for showering and hadronization.
For all of the non-resonant QCD backgrounds, one additional jet in the matrix element was allowed, and MLM matching~\cite{Mangano:2006rw, Alwall:2007fs} was implemented with the parameter {\tt xqcut} set to $30\gev$ to prevent over-counting in phase space.
For all of the non-resonant backgrounds, a common set of generator cuts was used: Jets produced in the hard process were required to have $p_{T,j} > 20 \gev$ and $|\eta_j| < 4.0$ while photons were required to have $p_{T,\gamma} > 25\gev$ and $|\eta_{\gamma}| < 2.7$. For backgrounds where a photon pair is produced in the hard process, the invariant mass of the photon pair was limited to $90 < m_{\gamma\gamma} < 180 \gev$. Finally, the invariant mass of a $b$-quark pair was required to be greater than $45\gev$, and all pairs of light jets (including $g$, $u$, $d$, $s$, and $c$) were required to have $m_{jj} > 25\gev$. To avoid overlaps between the different samples, events were vetoed in the $cc\gamma\gamma, jj\gamma\gamma$ and $ccj\gamma$ samples if they contained two $b$-quarks, and from the $jj\gamma\gamma$ samples if they contained two $c$-quarks.

Finally, the $t\bar{t}$ and $t\bar{t}\gamma$ processes were produced directly at LO in \MadGraph\ interfaced to \Pythia\ with the CTEQ6L1 PDF set. The $t\bar{t}$ total cross section at NNLO in QCD was recomputed at both $14$ and $27\tev$, with soft-gluon resummation effects to NNLL, using {\sc\small Top++2.0}\cite{Czakon:2011xx}. The NNLO+NNLL cross section is then used to normalize the events. Events in the $t\bar{t}$ sample were vetoed if they contained any true photons. For $t\bar{t}\gamma$, the NLO cross section is available at $14\tev$~\cite{Melnikov:2011ta}. We use this cross section to normalize our $14\tev$ sample, and compute a $k$-factor, $k = 1.48$, which we take to be the same at $27\tev$.

\subsubsection{Details of the $ggF$ Background}\label{ss.ggf_background}

The background from single-Higgs production via gluon fusion turns out to be a dominant contribution to the background for $hh$ production, as the true $h\to\gamma\gamma$ process is irreducible, and QCD interactions lead to a significant continuum of $b\bar{b}$, some of which will be indistinguishable from the $h\to b\bar{b}$ signal regardless of cuts. However, because the lowest-order process is lacking the two partons necessary to fake a $b\bar{b}$ signal, the background is dependent on different modeling choices. To account for all potential production mechanisms of the additional partons, we generate the sample inclusively in the heavy top-mass limit, allowing up to two extra partons in the matrix element, where the partons can be any light jets or a $c$- or $b$-quark. This is done with \MadGraph, with all other generator details the same as for the non-resonant QCD backgrounds. As with the signal process, the Higgs in the sample is decayed to two photons using \MadSpin\ before passing events to \Pythia\ and \Delphes\, with MLM matching between the parton-level events and the parton shower.  
Note that this is different than how this process was simulated in~\cite{Chang:2018uwu} where this background was generated with  \PowhegBox, interfaced to \PythiaSix.  In~\cite{Chang:2018uwu} this was done to mimic the ATLAS background generation of this process for the HL-LHC~\cite{ATL-PHYS-PUB-2017-001}, which we also compare to for validation in Appendix~A.  We have found roughly similar results with  \PowhegBox\,and \MadGraph\, however, the  \PowhegBox\ codes generate separately at NLO the $h$, $h+j$, and $h+jj$ processes, and in~\cite{Chang:2018uwu} only the $h$ code was used, which in turn means the additional $b$-jets are generated only from the parton shower, or splitting from the single emission interfaced to the parton shower.  The double real emission graph is not included in the method of ATLAS~\cite{ATL-PHYS-PUB-2017-001} and~\cite{Chang:2018uwu}, nor is the hard gluon splitting diagram. We find it is much simpler to generate the inclusive matched sample in \MadGraph\ in the infinite top mass limit (which \PowhegBox\, also takes) with all the relevant processes which lead to hard $b$'s. 
While real emission terms $\propto \alpha_S^2\,y_b^2$ are included in the $b\bar{b}H$ background in Refs.~\cite{ATL-PHYS-PUB-2017-001, Chang:2018uwu} via \Pythia, these include only the terms $\propto \alpha_S^2\,y_b^2$, neglecting the $\alpha_S^3\,y_b\,y_t$ and $\alpha_S^4\,y_t^2$ terms included in our method. The $\alpha_S^3\,y_b\,y_t$ terms were considered in Ref.~\cite{Wiesemann:2014ioa}, while the $\alpha_S^4\,y_t^2$ terms have only recently been considered at NLO in Ref.~\cite{Deutschmann:2018avk}.
We comment on these differences further in both section \ssref{ggF_results} and Appendix A, and demonstrate the robustness of our estimate. For a comparison of $b\bar{b}h$ production between different generators, but which do not include the $\alpha_s^4\,y_t^2$ contribution, see Ref.~\cite{DeFlorian:2016spz}.

Because our final selection will require two additional jets with respect to the LO process, a significant portion of the selected $ggF$ sample will be from the $h + 2j$ portion of the sample, so using the inclusive $k$-factor (known to N$^3$LO~\cite{Anastasiou:2015ema, Anastasiou:2016cez}) would be an overestimate. The $k$-factor for exclusive $h+2j$ production has been computed at NLO~\cite{Campbell:2006xx, Heinemeyer:2013tqa, Greiner:2015jha}, and was found to be a mild $\sim 15\%$ effect. While the $m_t \to \infty$ limit is known to be inaccurate for energies $\gtrsim 2 m_t$, the bulk of the background comes from events with $p_{T,h} \sim m_h$, where finite top-mass effects are expected to be small, as has been shown explicitly for exclusive $h+j$ production~\cite{Jones:2018hbb,Hirschi:2015iia}. Note also that our choice of $p_T$ cuts in \sref{results} also eliminates effects from the interference of top- and bottom-quark loops in the $h+j$ sample\footnote{See, e.g., Figure 2 of Ref.~\cite{Hirschi:2015iia} in the context of $h+j$. We expect the results for $h+2j$ to be similar.}. Additional cross-checks of our modeling of the $ggF$ background are described in \ssref{ggF_results}.

\section{HE-LHC measurement}
\label{s.calculation} \setcounter{equation}{0} \setcounter{footnote}{0}

An exciting possibility for the LHC is an upgrade to center of mass energies of $\sqrt{s} = 27\tev$. This could be achieved by upgrading the $8.33\,\mathrm{T}$ field dipole magnets currently installed at the LHC with $16\,\mathrm{T}$ dipole fields that are being developed for FCC-hh studies~\cite{Apollinari:2017}. Such an upgrade would substantially extend the reach of the LHC for new physics.

For the remainder of this study, we consider two benchmark scenarios: HL-LHC with $\sqrt{s} = 14\tev$ and an integrated luminosity $L = 3\,\mathrm{ab}^{-1}$ (the same as considered in~\cite{ATL-PHYS-PUB-2017-001}) and HE-LHC with $\sqrt{s} = 27\tev$ and $L = 15\,\mathrm{ab}^{-1}$, which represents the full improvement possible with an upgraded collider. An upgrade to $27\tev$ will present a number of challenges to future detectors, particularly in understanding the effects of pile-up interactions. For simplicity, we assume that the projected performance of the ATLAS detector at the HL-LHC~\cite{CERN-LHCC-2015-020, ATL-PHYS-PUB-2016-026} can be replicated. We also consider the effects of improved electromagnetic calorimeter resolution, which would help discriminate true $h\rightarrow \gamma\gamma$ decays from background. These two detector performance scenarios are discussed more in the following section.

\subsection{Detector Assumptions and Event Reconstruction}\label{ss.detector}

To approximate the HE-LHC detector scenario, we use \Delphes~\cite{DeFavereau:2013fsa} to simulate the performance of an upgraded ATLAS-like detector.
A custom input card based off the current ATLAS card available in \DelphesFull\ is used, with slight modifications to better match the projected performance of the ATLAS and CMS detectors for HL-LHC based on the current scoping documents~\cite{CERN-LHCC-2015-020, CMSCollaboration:2015zni}
and other projections~\cite{ATL-UPGRADE-PUB-2013-014, ATL-PHYS-PUB-2016-026, ATL-PHYS-PUB-2016-025}.
These modifications were validated by comparing to the most recent projection for measuring $\lambda_3$ at $14\tev$~\cite{ATL-PHYS-PUB-2017-001} (see \aref{validation} for details). The custom card also allows us to directly simulate the $b$-tagging performance, as well as the rates for jets or electrons to fake a photon in the detector, discussed in more detail below.

The efficiency of identifying photons is taken to match the results of the \doubleq{Tight-ID} requirement in~\cite{ATL-PHYS-PUB-2016-026}, which reaches approximately $85\%$, but falls off for photons with $p_T \lesssim 120\gev$.
The probability of an electron being misidentified as a photon is taken to be $2\%$ for $|\eta| < 1.7$ and $5\%$ otherwise~\cite{ATL-PHYS-PUB-2017-001}.

Following showering and hadronization, jets are clustered in \Delphes\ with the \FastJet\ package~\cite{Cacciari:2011ma} using the anti-$k_t$ algorithm~\cite{Cacciari:2008gp} with a distance parameter of $R = 0.4$. The $b$-tagging efficiencies and mis-tagging rates are taken as functions of $p_T$ as follows:
\begin{align}
  p_{b\rightarrow b} & = 0.70 \cdot \left(\frac{34.3 \tanh\left(p_T / 330\gev\right)}{1 + p_T / 11.6\gev}\right), \\
  p_{c\rightarrow b} & = 0.20 \cdot \left(\frac{28.6 \tanh\left(p_T/50\gev\right)}{1 + p_T / 290\gev}\right), \\
  p_{j\rightarrow b} & = 0.007,
\end{align}
where here only $j$ indicates a jet that contains no $b$ or $c$ quarks. These probabilities correspond to roughly $70\%,  20\%$, and $\lesssim 1\%$ respectively. The probability for a light jet to fake a photon in the detector is also taken as a function of $p_T$:
\begin{equation}
  p_{j\rightarrow \gamma} =
  \begin{cases}
	\left(\frac{p_{T,j}}{8\gev} - 2.5\right) \times 10^{-4}		& 30.0 < |p_{T,j}| \leq 60.0\gev \\
	\left(7\cdot \exp\left(\frac{-p_{T,j}}{100\gev}\right) + 1.1\right)\times {10^{-4}} 	& |p_{T,j}| > 60\gev,
\end{cases}
\end{equation}
which peaks at $5\times 10^{-4}$ for $p_{T,j} \sim 60\gev$ before dropping asymptotically to $1.1\times 10^{-4}$.

Aside from the usual jet energy scale correction applied in \Delphes\ at simulation level, an additional correction is made to the momentum of $b$-tagged jets at analysis level by scaling their four-momentum by a function of the jet $p_T$ that starts at $1.08$ for low $p_T$ and falls off smoothly to $1.0$ around $p_{T,j} \sim 100\gev$. This is to help correct for energy losses during reconstruction, and is a rough approximation to the {\em ptcorr} correction used by ATLAS~\cite{Aaboud:2017xsd}. The correction was found to give a $m_{b\bar{b}}$ peak centered at $m_h = 125\gev$, and slightly improve the resolution on the $b\bar{b}$ invariant mass (See \aref{validation} for more details).

\subsubsection{E-Cal Resolution}

To simulate the resolution of photons in the Electromagnetic Calorimeter, two parameterizations were used. The first parameterization (hereon referred to as \doubleq{Regular}, or \doubleq{Reg.}) was fixed to best fit the performance expected at HL-LHC based on~\cite{ATL-PHYS-PUB-2016-026}. The second parameterization is taken from the \doubleq{Medium} benchmark from $100\tev$ studies of double Higgs production~\cite{Contino:2016spe}, hereon referred to as \doubleq{Improved} or \doubleq{Imp.} The corresponding equations are:

\begin{align}
\text{Regular:}
& \qquad &  \frac{\sigma_{E}^{\text{ECal}}}{E}  & =
  \begin{cases}
  	 \frac{0.2233}{\sqrt{E}} \oplus \frac{0.974}{E},			& |\eta| \leq 1.52 \\
	 0.02 \oplus \frac{0.131}{\sqrt{E}} \oplus \frac{1.31}{E},	& 1.52 < |\eta| \leq 3.20 \\
	 0.0385 \oplus \frac{0.3135}{\sqrt{E}}					& |\eta| > 3.20
  \end{cases} \\
\text{Improved:}
& \qquad &  \frac{\sigma_E^{\text{ECal}}}{E}  & =
  	0.01 \oplus \frac{0.1}{\sqrt{E}}, 										 \\
\end{align}

These give a resolution of $\sim2.5\%$ and $1.4\%$ at $E = 100\gev$. In both scenarios, the hadronic calorimeter resolution was parameterized as
\begin{equation}
  \frac{\sigma_{E}^{\text{HCal}}}{E} =
  \begin{cases}
  	0.03 \oplus \frac{0.52}{\sqrt{E}} \oplus \frac{1.59}{E} 		& |\eta| \leq 1.7 \\
	0.05 \oplus \frac{0.71}{\sqrt{E}}						& 1.7 < |\eta| \leq 3.20. \\
  \end{cases}
\end{equation}

The \doubleq{Regular} parameterization is used to validate our setup with previous ATLAS studies at $14\tev$, as described in \aref{validation}.

\section{Results and Comparison to Other Colliders}
\label{s.results} \setcounter{equation}{0} \setcounter{footnote}{0}

\subsection{Event Selection}

For tabulating the signal and backgrounds, we consider isolated photons and jets with $p_{T} > 30\gev$ and $|\eta| < 2.5$. A reweighting procedure is then implemented to account for the probabilities associated with $b$-tagging and light jets faking photons listed above. We select events containing at least two photon and $b$-jet candidates to be consistent with the signal. The weighted events are then subjected to selection criteria chosen to optimize the SM di-Higgs signal strength.

The leading (sub-leading) photon and $b$-jet are required to have $p_T > 60~(35)\gev$. The diphoton invariant mass is required to satisfy $122 < m_{\gamma\gamma} < 128\gev$, ($122.8 < m_{\gamma\gamma} < 127.2\gev$) in the \doubleq{Regular} (\doubleq{Improved}) scenario, while the invariant mass of the $b$-quark pair is constrained to $100 < m_{b\bar{b}} < 150\gev$ in both cases.

Compared to the ATLAS $14\tev$ study~\cite{ATL-PHYS-PUB-2017-001}, we impose somewhat stricter cuts on the $p_T$ of each reconstructed Higgs, demanding that $p_{T,h} > 125\gev$. Our requirements for the final state particle $p_T$ are also somewhat more stringent. The cuts on the invariant radial distance between each jet or photon in a reconstructed Higgs are instead looser, requiring only $\Delta R_{b\bar{b}}, \Delta R_{\gamma\gamma} < 3.5$. The cuts on $p_T$ and $\Delta R$ of the reconstructed Higgs are tightly correlated with cuts on the invariant mass of the Higgs pair, $m_{hh}$, and these values correspond roughly to requiring $m_{hh} \gtrsim 350\gev$ (this is similar to the cut considered in \cite{Goncalves:2018yva}). We also impose an additional cut on the decay angle of the Higgs boson pair evaluated in the lab frame, requiring $|\cos\theta_{hh}| < 0.8$. This was seen in refs.~\cite{Yao:2013ika, He:2015spf} to significantly reduce backgrounds from QCD processes as well as $t\bar{t}h$ production
As in ref.~\cite{ATL-PHYS-PUB-2017-001}, events with any isolated leptons and more than 5 jets are rejected to reduce backgrounds from top quark decays.\\

\begin{figure}[!htbp]
\centering
\begin{subfigure}{0.49\linewidth}
  \centering
  \includegraphics[width=\linewidth]{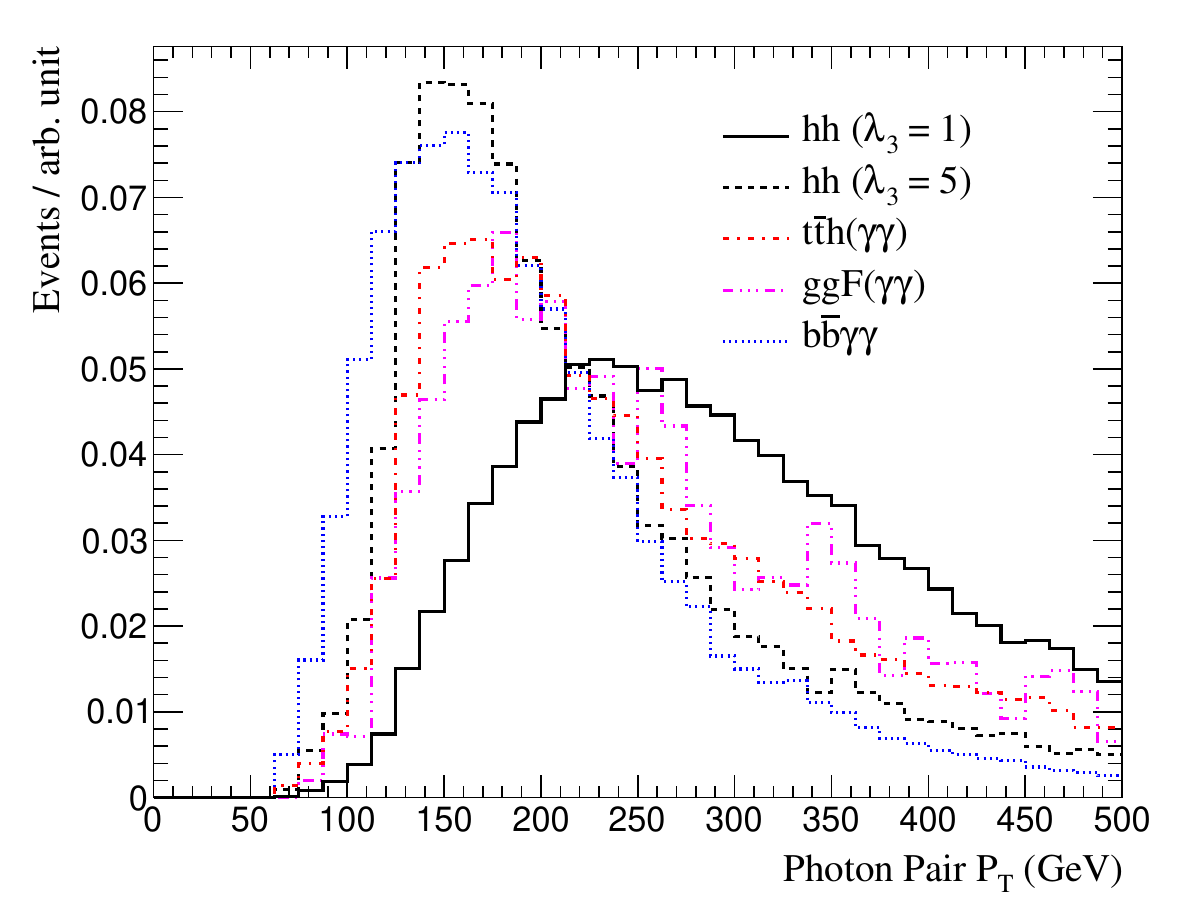}\caption{}
\end{subfigure}
\hfill
\begin{subfigure}{0.49\linewidth}
  \centering
  \includegraphics[width=\linewidth]{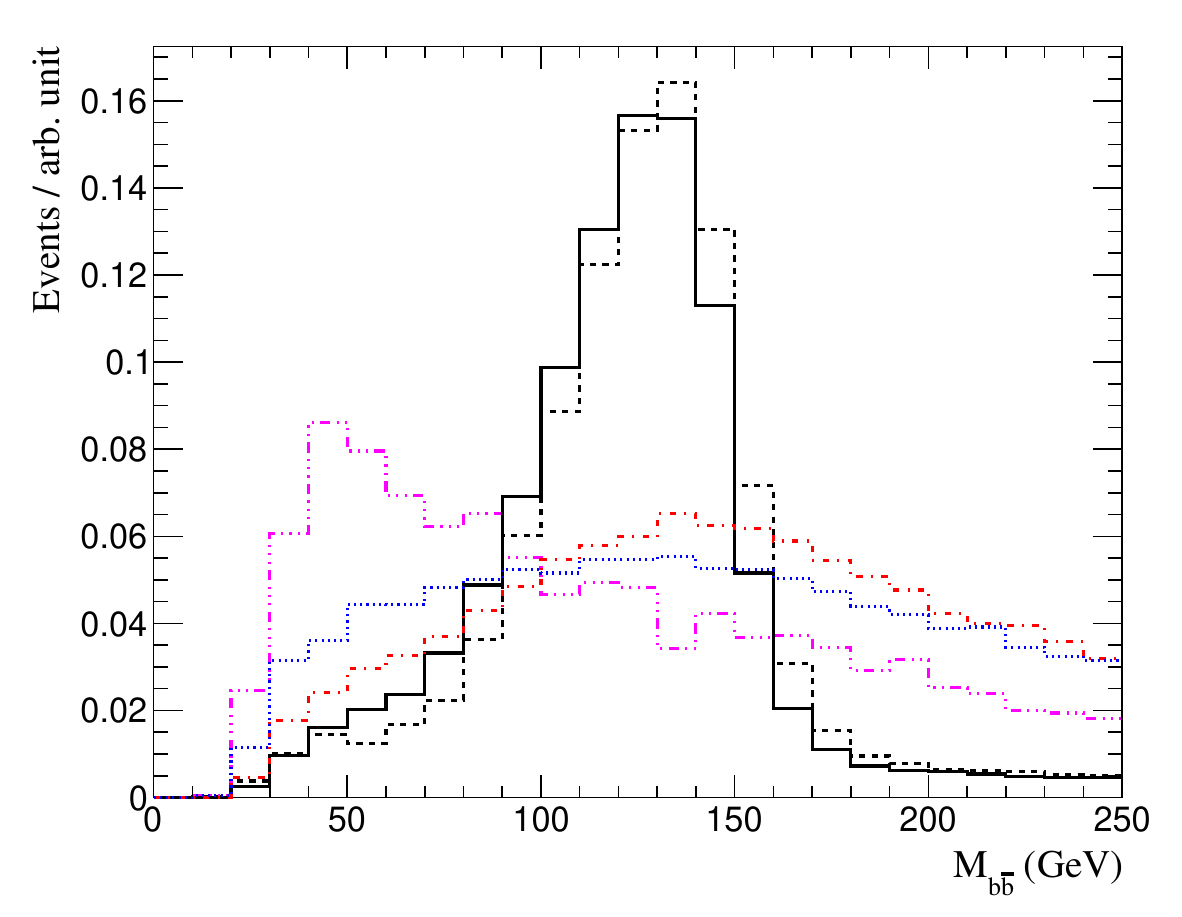}\caption{}
\end{subfigure}
    \\[\baselineskip]
\begin{subfigure}{0.49\linewidth}
  \centering
  \includegraphics[width=\linewidth]{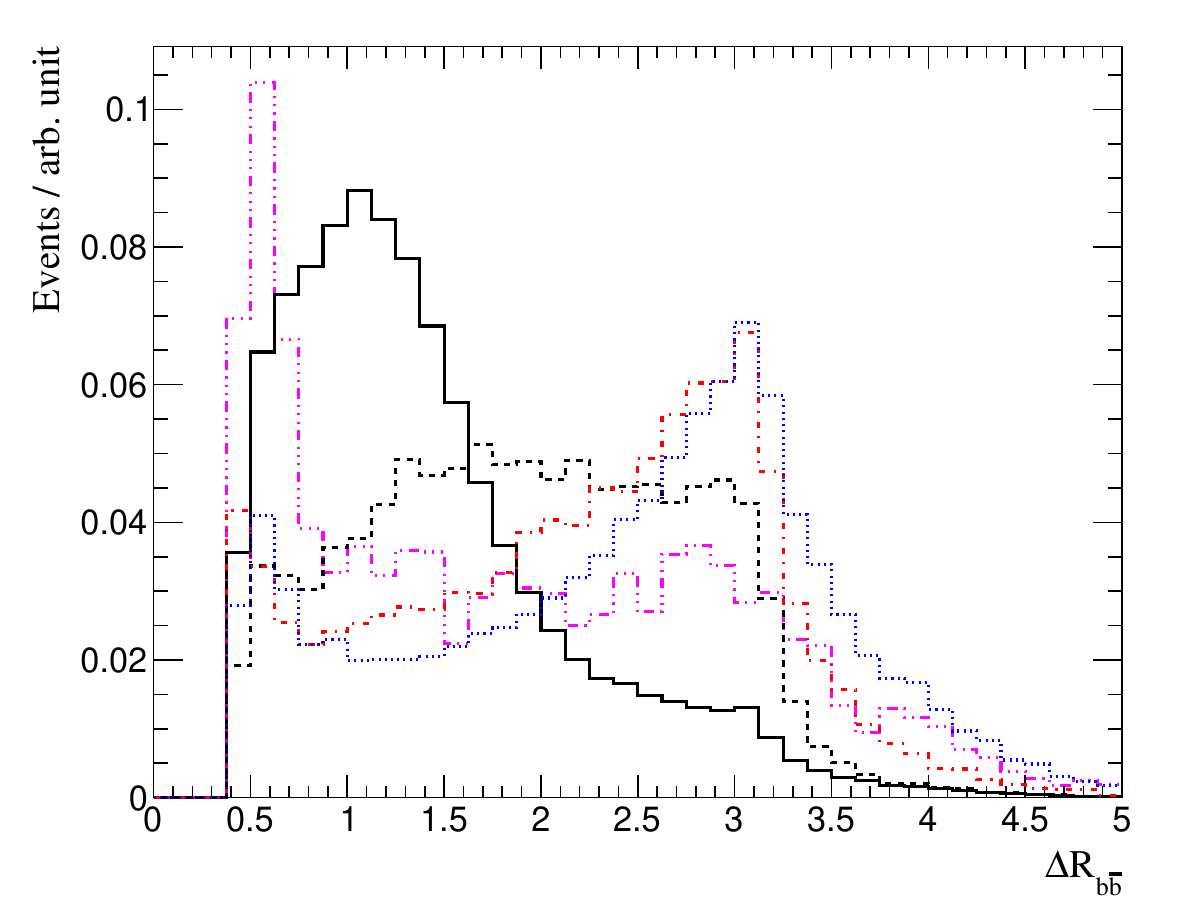}\caption{}
\end{subfigure}
\hfill
\begin{subfigure}{0.49\linewidth}
  \centering
  \includegraphics[width=\linewidth]{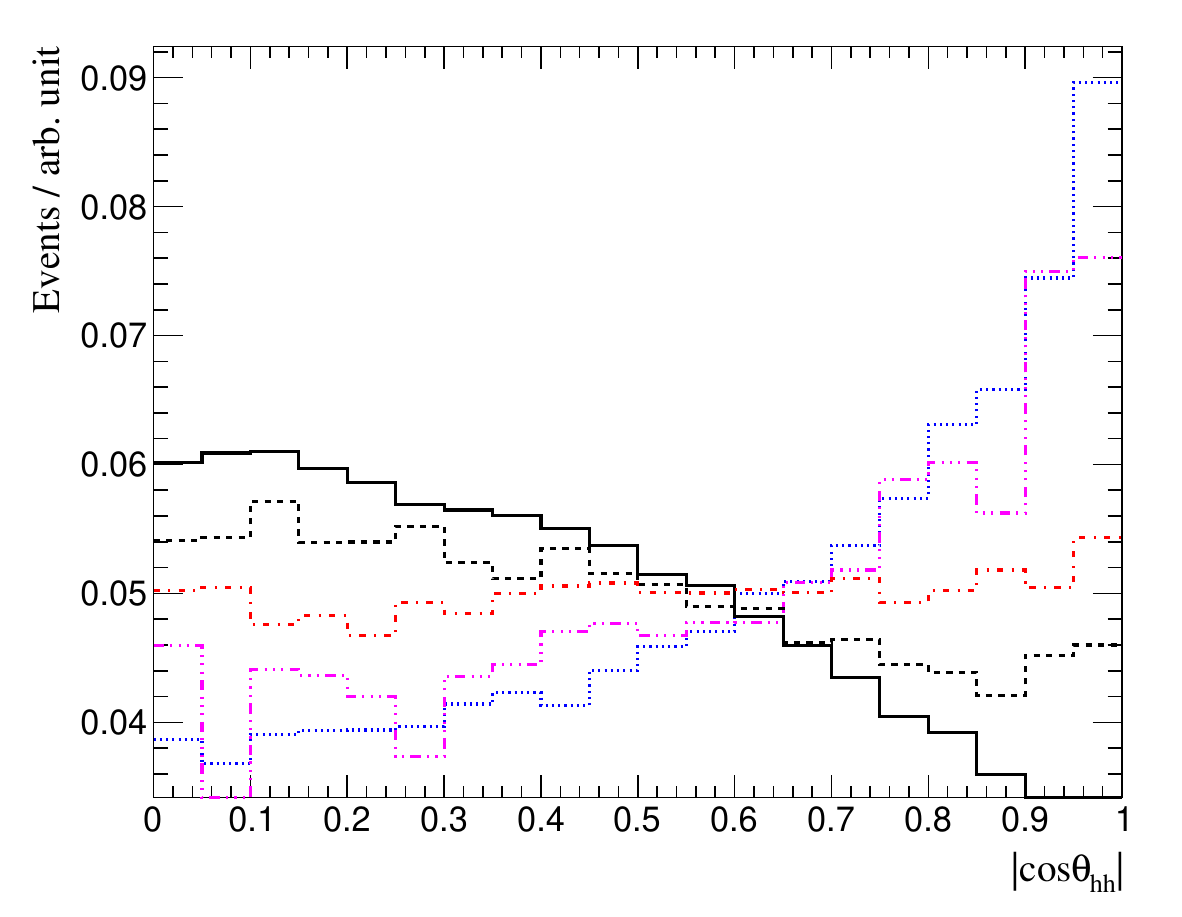}\caption{}
\end{subfigure}
\caption{Normalized distributions of (a) the $p_T$ of the reconstructed $h\rightarrow\gamma\gamma$, (b) the invariant mass of the $h\rightarrow b\bar{b}$ pair, (c) angular separation ($\Delta R_{b\bar{b}}$) of the $b$-quark pair, and (d) the magnitude of $\cos\theta_h$, the Higgs decay angle (see text for details). We show the distributions for the SM ($\lambda_3$ = 1) signal $hh\rightarrow b\bar{b}\gamma\gamma$ (black, solid), as well as the signal when $\lambda_3 = 5$ (green, dot-dashed), and the primary backgrounds: $t\bar{t}h(\gamma\gamma)$ (red, dashed), $Zh(\gamma\gamma)$ (pink, dot-dot-dashed), and the irreducible $b\bar{b}\gamma\gamma$ background from QCD (blue, dotted) which is representative of the other QCD backgrounds. For all distributions we plot only the events where two photons and two $b$-jets have been reconstructed, satisfying the final $p_T$ and $\eta$ cuts, and with no isolated leptons.}
\label{fig:kinematics}
\end{figure}

The full list of analysis cuts is summarized below:

\begin{itemize}
  \item At least 2 isolated photons and $b$-tagged jets with leading $p_T > 60\gev$ and subleading $p_T > 35\gev$, all with $|\eta_{\gamma,b}| < 2.5$.
  \item $p_{T,\gamma\gamma}, p_{T,b\bar{b}} > 125\gev$.
  \item $\Delta R_{b\bar{b}}, \Delta R_{\gamma\gamma} < 3.5$
  \item $|m_{\gamma\gamma} - 125.0\gev| <
  		\begin{cases}
			3.0\gev 	& \text{Reg.}\\
			2.2\gev 	& \text{Imp.}
		\end{cases}$
  \item $|m_{b\bar{b}} - 125.0\gev| < 25\gev$.
  \item $n_{\text{jets}} < 6$ for jets with $p_T > 30\gev$, $|\eta| < 2.5$.
  \item No isolated leptons with $p_T > 25\gev$.
  \item $|\cos\theta_{hh}| < 0.8$.
\end{itemize}

\begin{figure}[!htbp]
\centering
\includegraphics[width=0.65\linewidth]{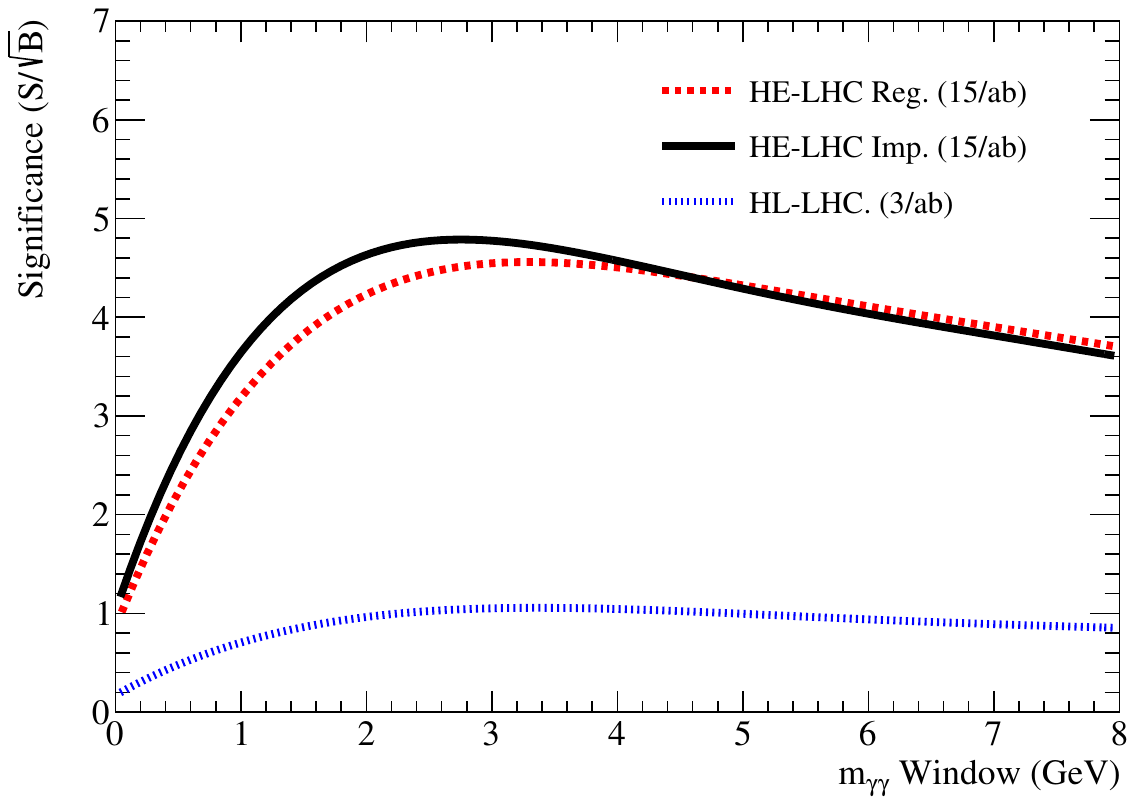}
\caption{Plot of the expected significance ($S/\sqrt{B}$) for each set of signal and background samples as a function of the window on $m_{\gamma\gamma}$ ($|m_{\gamma\gamma} - 125.0 \, {\rm GeV} | < m_{\gamma\gamma} {\textrm{Window}}$) on the $h\rightarrow\gamma\gamma$ peak. The peak of each curve is used to determine the optimal selection cut in each scenario.}
\label{fig:sig_v_window}
\end{figure}

To optimize the cut on $m_{\gamma\gamma}$, the expected significance ($S/\sqrt{B}$) was computed for each set of samples with the width of the $m_{\gamma\gamma}$ window around $m_h = 125\gev$ varied up to $8 \gev$. The results are plotted in \figref{sig_v_window}.

The selection efficiency for the di-Higgs signal is $2.56\%$ for both the Reg. and Imp. scenarios. Note that, as can be seen in Figs. \ref{fig:higgs_parton} and \ref{fig:kinematics}, the kinematics of the $hh\rightarrow b\bar{b}\gamma\gamma$ signal change for non-SM values of $\lambda_3$, and the optimal selection criteria for the SM signal are not necessarily the same as for other values. For our purposes, we assume that any deviation from $\lambda_3 = 1$ is small, and leave an analysis optimized for excluding significant departures from the SM to future work.

\subsection{Expected Event Yields}

The expected number of events from each signal and background channel based on our simulations is shown in \tref{results} assuming $3\,\mathrm{ab}^{-1}$ ($15\,\mathrm{ab}^{-1}$) integrated luminosity for $14\tev$ ($27\tev$). The uncertainty for each sample is estimated by partitioning the full MC sample into subsamples and computing the standard deviation of the results from each subsample.

\begingroup
\renewcommand*{\arraystretch}{1.1}
\begin{table}[ht]
\centering
\begin{tabular}{|c|
  rcl
  rcl
  rcl|}
\hline
\multicolumn{1}{|c|}{\multirow{3}{*}{Process}}  & \multicolumn{9}{c|}{Expected Events} \\ \cline{2-10}
\multicolumn{1}{|c|}{}	& \multicolumn{3}{c|}{\multirow{2}{*}{$14\tev$ ($3\,\mathrm{ab}^{-1}$)}} 	& \multicolumn{6}{c|}{$27\tev$ ($15\,\mathrm{ab}^{-1}$)}\\
\multicolumn{1}{|c|}{}	& \multicolumn{3}{c|}{}									& \multicolumn{3}{c}{Reg.} & \multicolumn{3}{c|}{Imp.} \\ \hline
$h(b\bar{b})h(\gamma\gamma)$	& $9.35$	& $\pm $ & $0.03$	& $157.4$  & $\pm$ & $0.2$	& $157.4$	& $\pm$ & $0.4$ \\
\hline
$t\bar{t}h(\gamma\gamma)$    	& $7.82$   & $\pm $ & $0.06$	& $157.6$	& $\pm$ & $1.3$	& $153.4$	& $\pm$ & $0.6$ \\
$Zh(\gamma\gamma)$           	& $3.88$   & $\pm $ & $0.05$	& $53.5$  	& $\pm$ & $0.9$	& $63.0$	& $\pm$ & $0.6$ \\
$ggF(\gamma\gamma)$          	& $11.4$   & $\pm $ & $0.5$	& $240.0$	& $\pm$ & $6.5$	& $219.6$	& $\pm$ & $5.8$ \\
$b\bar{b}\gamma\gamma$       	& $27.3$  	& $\pm $ & $1.7$	& $152.7$	& $\pm$ & $2.2$	& $126.0$	& $\pm$ & $11.9$ \\
$c\bar{c}\gamma\gamma$       	& $4.8$   	& $\pm $ & $0.3$	& $80.6$  	& $\pm$ & $4.4$	& $57.4$	& $\pm$ & $1.4$ \\
$jj\gamma\gamma$             	& $2.3$   	& $\pm $ & $0.2$	& $40.6$  	& $\pm$ & $3.3$	& $16.9$	& $\pm$ & $0.3$ \\
$b\bar{b}j\gamma$            	& $7.3$   	& $\pm $ & $1.9$	& $164.5$	& $\pm$ & $9.0$	& $177.8$	& $\pm$ & $9.5$ \\
$c\bar{c}j\gamma$            		& $1.5$   	& $\pm $ & $0.4$	& $35.5$  	& $\pm$ & $3.1$	& $34.4$	& $\pm$ & $6.8$ \\
$b\bar{b}jj$                 		& $6.5$   	& $\pm $ & $0.5$	& $166.9$	& $\pm$ & $8.5$	& $152.7$	& $\pm$ & $6.9$ \\
$Z(b\bar{b})\gamma\gamma$   & $1.4$   	& $\pm $ & $0.1$	& $12.3$  	& $\pm$ & $0.4$	& $10.6$	& $\pm$ & $0.4$ \\
$t\bar{t}$               			& $1.3$  	& $\pm $ & $0.3$	& $16.9$  	& $\pm$ & $2.0$	& $4.1$	& $\pm$ & $0.7$ \\
$t\bar{t}\gamma$            		& $3.6$   	& $\pm $ & $0.5$	& $95.1$	& $\pm$ & $7.7$	& $101.5$	& $\pm$ & $8.0$ \\
\hline
Total Background			& $78.9$  	& $\pm $ & $1.8$	& $1216$	& $\pm$ & $19$	& $1117$	& $\pm$ & $20$ \\ \hline
Significance ($S/\sqrt{B}$)		& $1.05$  	& $\pm $ & $0.01$	& $4.51$  	& $\pm$ & $0.04$	& $4.70$	& $\pm$ & $0.04$ \\ \hline
\end{tabular}
\caption{The expected number of events from each signal/background process at $14$ and $27\, {\rm TeV}$ based on MC simulation, using the cross sections shown in \tref{background_csx}. Also shown are the total background and significance (computed as $S/\sqrt{B}$) at each energy.}
\label{t.results}
\end{table}
\endgroup

The expected significance at 14\tev\ (computed as $S/\sqrt{B}$) is found to be $1.05 \pm 0.01$, comparable to previous projections~\cite{ATL-PHYS-PUB-2017-001}. At $27\tev$, with $15\,\mathrm{ab}^{-1}$, a significance of $4.51 \pm 0.04$  ($4.70 \pm 0.04$) is attainable in the Regular (Improved) scenario.

\begin{figure}[!htbp]
\centering
\begin{subfigure}{0.49\textwidth}
  \centering
  \includegraphics[width=\linewidth]{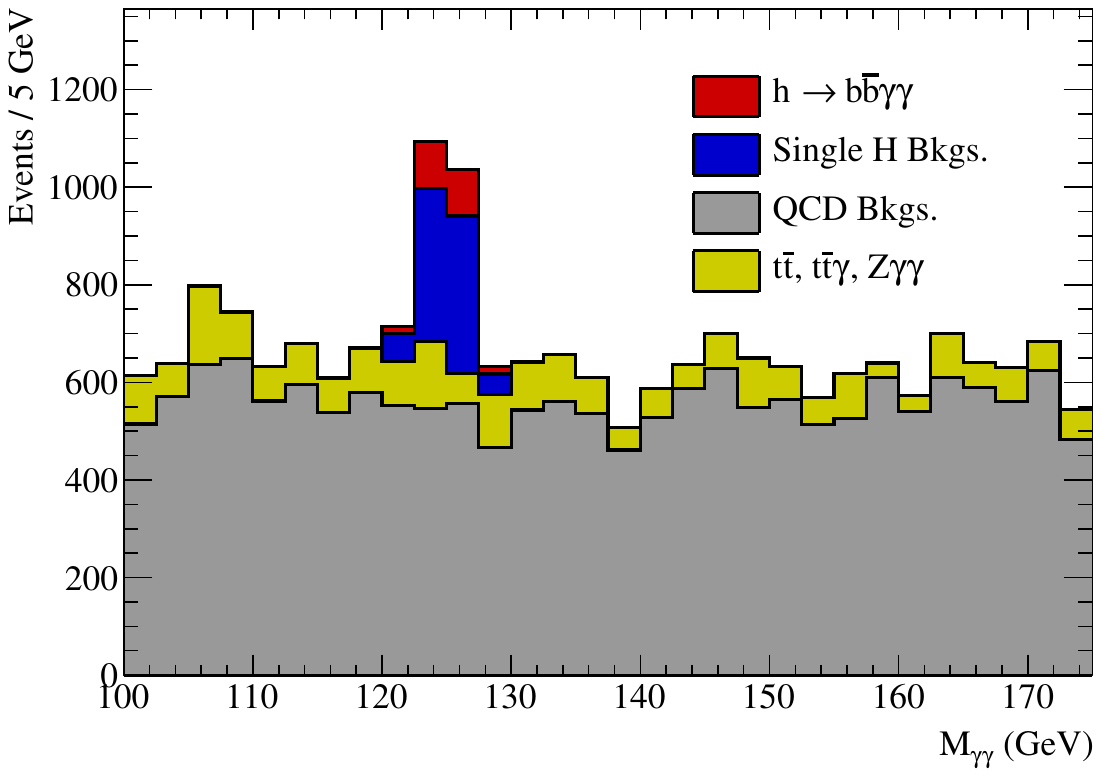}
  \caption{}
  \label{fig:stacked_gamma}
\end{subfigure}
\begin{subfigure}{0.49\textwidth}
  \centering
  \includegraphics[width=\linewidth]{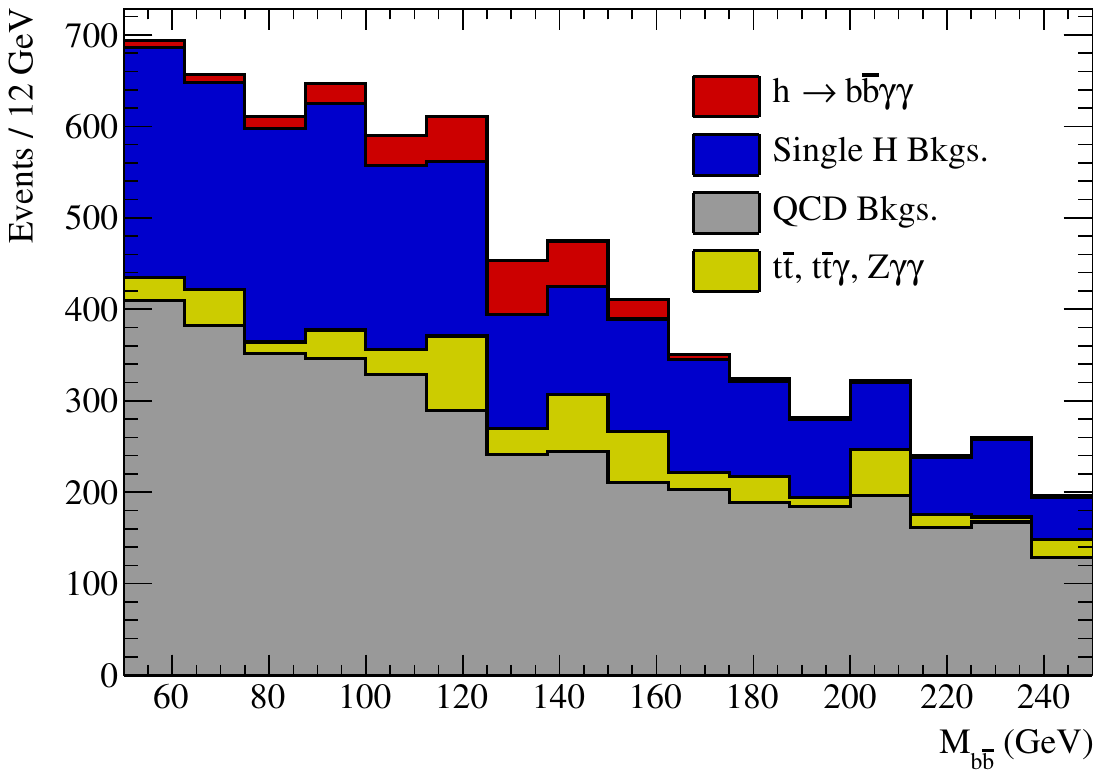}
  \caption{}
  \label{fig:stacked_bb}
\end{subfigure}
\caption{Stacked histograms showing the contribution of events from the signal (red) and each background category (single Higgs production in blue, $t\bar{t}$, $t\bar{t}\gamma$ and $Z\gamma\gamma$ in yellow, and QCD backgrounds in grey) passing every cut except the cut on $m_{\gamma\gamma}$ (top) and $m_{b\bar{b}}$ (bottom) in the \doubleq{Improved} detector scenario. The single Higgs background shows an additional peak at $m_{b\bar{b}} < 120\, {\rm GeV}$ due to the contribution from the $Z(b\bar{b})h(\gamma\gamma)$ background.}
\label{fig:stacked}
\end{figure}

As evident in \tref{results}, the most significant backgrounds arise from $ggF(\gamma\gamma)$, $t\bar{t}h(\gamma\gamma)$, $b\bar{b}\gamma\gamma$, and $b\bar{b}j\gamma$. The $ggF$ and $t\bar{t}h$ backgrounds are difficult to suppress, as they include a true $h\rightarrow \gamma\gamma$ decay that is indistinguishable from that of the $hh$ system, as demonstrated in \figref{stacked}. The $b\bar{b}\gamma\gamma$ background similarly can only be reduced by improving the calorimeter resolution and tightening the cut on the diphoton invariant mass. The $b\bar{b}j\gamma$ background, on the other hand, is strongly dependent on the jet-faking-photon probability, and could be significantly reduced if true photons can be more reliably distinguished from jets.

\subsubsection{The $ggF$ Background}\label{ss.ggF_results}

As shown in \tref{results}, we find the background from single-Higgs production via gluon fusion to be very significant, with an event rate of the same order as the signal.
This result differs from some previous estimates~\cite{ATL-PHYS-PUB-2017-001, ATL-PHYS-PUB-2014-019} which found the gluon fusion background to be small compared to the signal,
as well as other studies which considered it to be negligible compared to the other single Higgs and irreducible backgrounds~\cite{Yao:2013ika, Contino:2016spe, He:2015spf, Azatov:2015oxa, Barr:2014sga, Goncalves:2018yva}.
Ref.~\cite{Chang:2018uwu} does include the gluon-fusion background, in a similar setup as the ATLAS studies, and finds it to be $\sim 2$ times larger than previous estimates, but still smaller than our result shown above.

To understand the source of this large background, we perform a further analysis of the simulated events in our $ggF(\gamma\gamma)$ sample described above. While there are contributions from mistagged jets, we find that $\sim 75\%$ of the events passing all selection criteria arise from two true $b$-jets.
While the showering of events can lead to additional $b$-jets in our event samples, it's apparent that the production of a hard $b\bar{b}$ pair accompanying Higgs production via gluon-fusion is a non-negligible component of the background.

To understand this more fully, and as an extra check on the size of the $ggF(\gamma\gamma)$ background estimate, we compute the parton-level cross section for $gg \to hb\bar{b}$ production in the $m_t \to \infty$ limit, with $y_b$ set to zero to ensure we are including only the effects of a hard gluon splitting to $b\bar{b}$ in association with $gg\to h$. This is done with a set of generator level cuts intended to mimic the analysis cuts described above. In particular, we demand: 

\begin{itemize}
  \item $p_{T,b1} > 60\gev$, $p_{T,b2} > 35\gev$,
  \item $|\eta_b| < 2.5$,
  \item $0.4 < |\Delta R_{b\bar{b}}| < 3.5$,
  \item $m_{b\bar{b}} > 100\gev$,
  \item $p_{T,h} > 125\gev$\footnote{At parton level, this automatically fixes $p_{T,b\bar{b}} > 125\,\mathrm{GeV}$ as well.}.
\end{itemize}

The remaining cuts apply only to the true $h\to\gamma\gamma$ decay, and are expected to have an efficiency close to unity. After folding in the $h\to\gamma\gamma$ branching ratio, the resulting cross section is $4.9\times 10^{-2}\,\mathrm{fb}$ at $\sqrt{s} = 27\tev$, or $\sim 740$ events at $15\,\mathrm{ab}^{-1}$. While the requirement of two $b$-tags and other detector efficiencies will reduce this number, this represents only the leading process in $ggF(\gamma\gamma)$ and its contribution is clearly non-negligible, supporting the findings in \tref{results}.

Recently, the gluon fusion contributions to $b\bar{b}h$ production (which we include as part of the $ggF$ background) have been considered in more detail in Ref.~\cite{Deutschmann:2018avk}, including all terms in a power counting up to $\alpha_S^5 y_t^2$ (formally N${}^3$LO) in the heavy top-mass limit.
The results of Ref.~\cite{Deutschmann:2018avk} demonstrate that the $y_t^2$ terms dominate the cross section, particularly when $p_{T,h} > 125\,{\mathrm{GeV}}$ as in our final selection. This is in agreement with our findings above, and further demonstrates that the production of a Higgs in association with a $b\bar{b}$ pair constitutes a significant background to di-Higgs production searches.
Furthermore, Ref.~\cite{Deutschmann:2018avk} also considers QCD corrections to $b\bar{b}h$ production from gluon-fusion and finds that these increase the cross section by a factor of $\gtrsim 2$ in all relevant regions of phase space. Because our $ggF$ sample also includes other processes leading to a reconstructed $bb\gamma\gamma$ final state we do not include these corrections, but we note that they could in principle reduce our final projected significance by up to $0.5$.

\subsection{Limit Setting on the Higgs Self-Coupling}

\begin{figure}[!htbp]
  \centering
  \includegraphics[width=12cm]{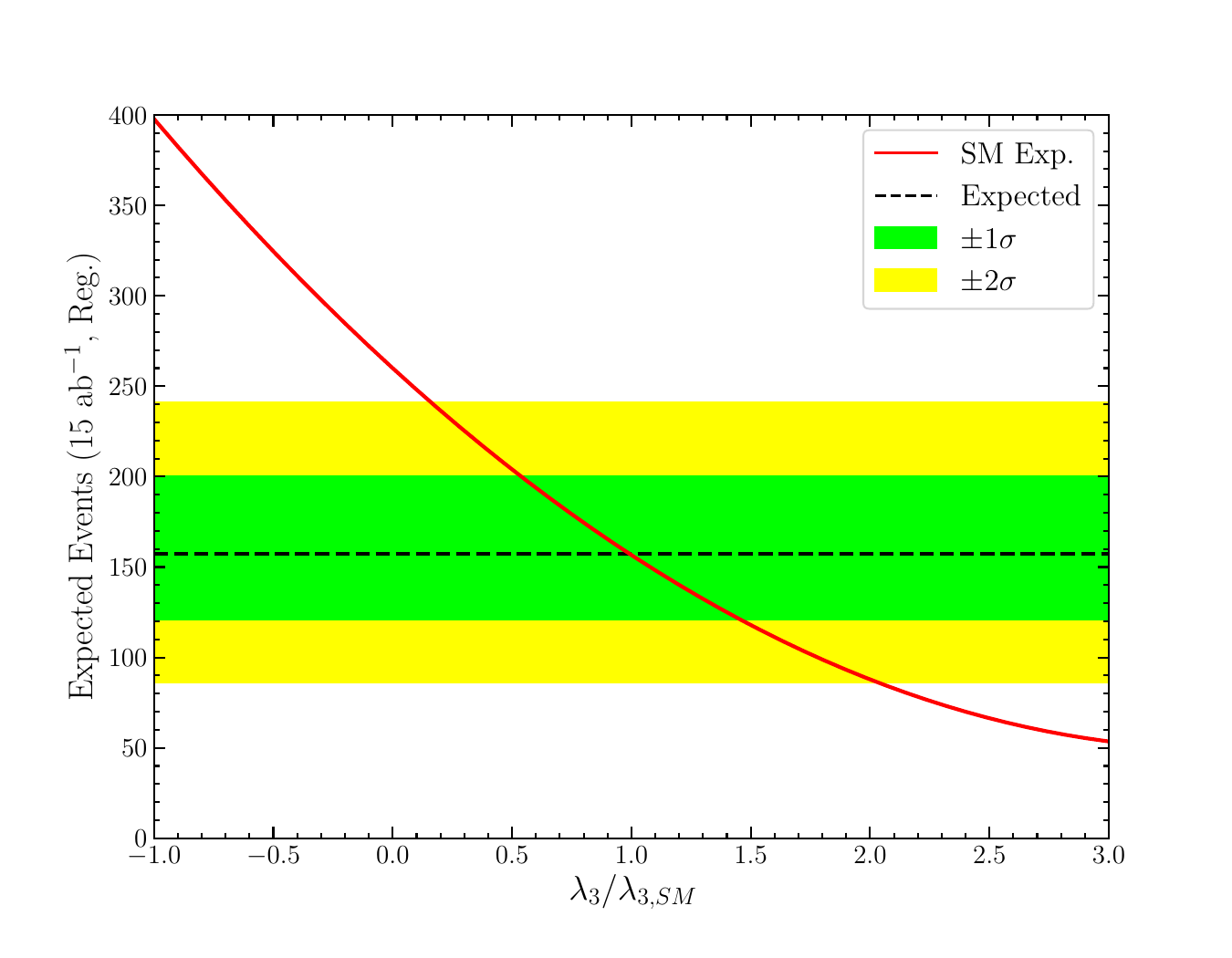}
  \caption{The expected number of signal events in a hypothetical experiment assuming the signal and background rates computed in \tref{results} at $L = 15\,\mathrm{ab}^{-1}$ for HE-LHC with the regular detector performance assumption. The black dashed line indicates the expected number of events from signal while the green (yellow) regions show the $1\sigma~(2\sigma)$ uncertainty regions arising from a likelihood scan with the statistical and MC uncertainties on the signal and background counts. The red curve shows the expected number of events from signal in a background free measurement as a function of $\lambda_3$, accounting for the changes in the signal acceptance due to kinematic differences at different $\lambda_3$.}\label{fig:lam3_limit_reg}
 \end{figure}

To understand the attainable precision on $\lambda_3$, we assume a hypothetical observation of $S + B$ events after all selection cuts, with $S$ and $B$ as given in \tref{results}. This allows us to derive $68$ and $95\%$ confidence intervals on the expected number of signal events using a likelihood scan, including only the MC and statistical uncertainties. The expected number of signal events with $15\,\mathrm{ab}^{-1}$ integrated luminosity is plotted in \figrefs{lam3_limit_reg}{lam3_limit_imp}, along with the $1\sigma~(2\sigma)$ regions in green (yellow).

We can also compute the expected number of events at a given integrated luminosity as a function of $\lambda_3$, taking into account both the varying $\sigma_{hh}$ cross section and the modified acceptance due to changes in the signal kinematics. For $L = 15\,\mathrm{ab}^{-1}$ this is shown in red in Figs. \figrefs{lam3_limit_reg}{lam3_limit_imp}. The intersection of this curve with the $1$ and $2\sigma$ regions indicate the expected precision on $\lambda_3$ (ignoring systematic uncertainties). We find

\begin{equation}
\lambda_3 \in
\begin{array}{c}	\left[0.54, 1.46\right] \\ \left[0.60, 1.46\right] \end{array}
\text{at} \quad 68\%\,\mathrm{C.L.} \quad
\begin{array}{c} \text{Reg.} \\ \text{Imp.} \end{array}
\end{equation}

Note that, as a result of the destructive interference discussed in \sref{intro}, there is a degeneracy in the expected number of events around $\lambda_3 \sim 5$. However, the kinematic structure of the $hh$ signal is very different at large values of $\lambda_3$, and such values could be easily rejected using differential measurements (e.g, with $m_{hh} = m_{b\bar{b}\gamma\gamma}$ or $p_{T,hh}$), so the degeneracy can be safely ignored for the purposes of this work.

 \begin{figure}[!htbp]
 	\centering
	\includegraphics[width=12cm]{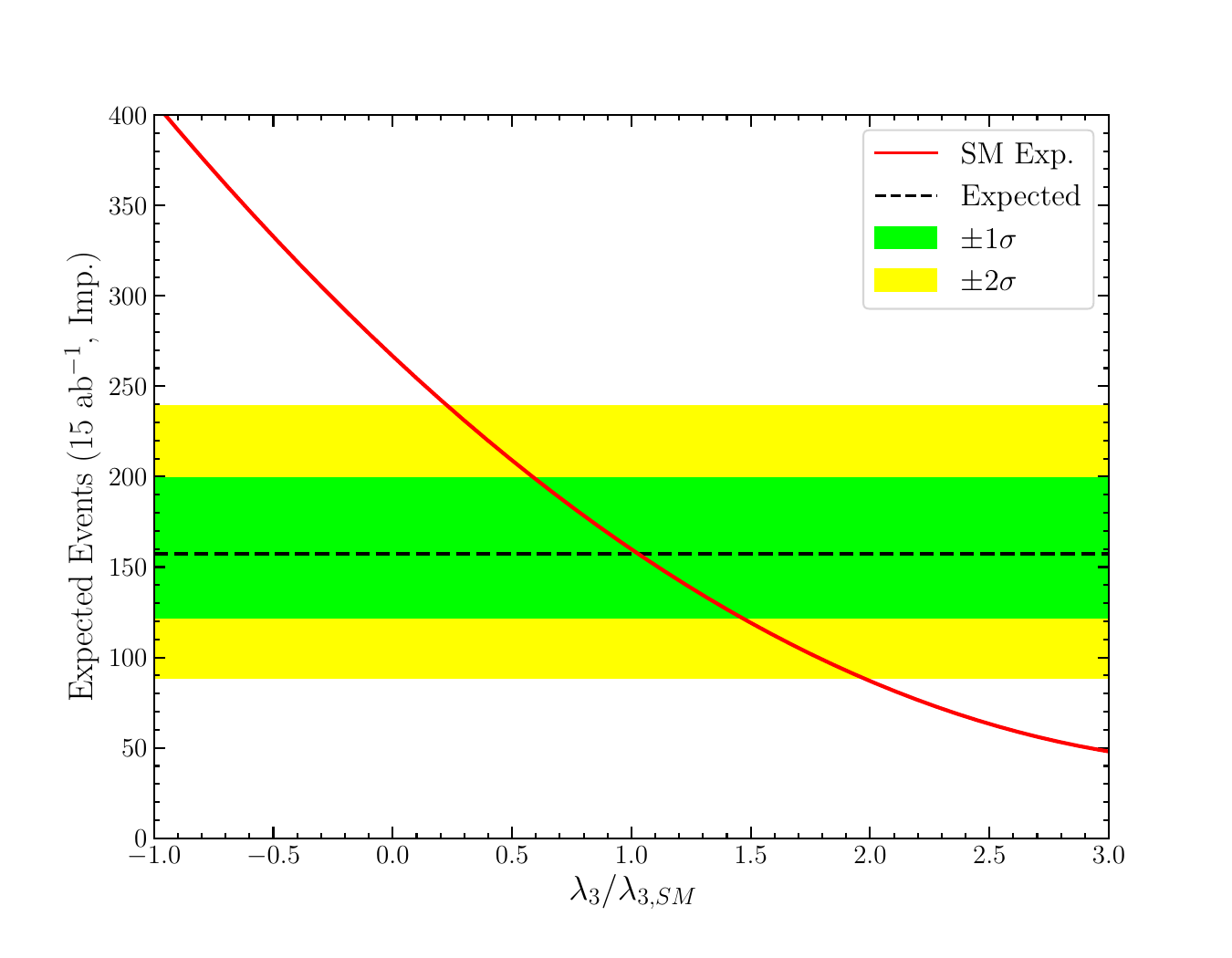}
	\caption{The same as \figref{lam3_limit_reg} but for the \doubleq{Improved} detector scenario.}\label{fig:lam3_limit_imp}
\end{figure}

\section{Conclusions}
\label{s.conclusions} \setcounter{equation}{0} \setcounter{footnote}{0}

We have presented a detailed study of the potential for a $\sqrt{s} = 27\tev$ hadron collider such as HE-LHC to search for di-Higgs production in the $b\bar{b}\gamma\gamma$ final state. In addition to the signal, all important backgrounds are simulated using Monte Carlo, with parton-showering and hadronization.
We consider detector scenarios based on the projected performance of the ATLAS detector at HL-LHC, and estimate the effects of the detector performance using the  \Delphes\ package. Backgrounds arising from the mistagging of a light flavor jet as a $b$-jet and from jets faking photons in the calorimeter have also been included using a reweighting procedure.  Compared to the study of~\cite{Goncalves:2018yva}, which also considered measuring $\lambda_3$ at a $27\tev$ collider, we find that the most important background comes from gluon fusion production of a single Higgs in association with additional $b$-jets. 
Recent studies \cite{Deutschmann:2018avk} indicate that the QCD corrections to this background may be very large, and our results make clear that further scrutiny of this background in all di-Higgs production searches is necessary.
This work complements the results of~\cite{Goncalves:2018yva} by considering the additional backgrounds arising from the aforementioned $ggF(\gamma\gamma)$, as well as $Z\gamma\gamma$, $t\bar{t}$, $t\bar{t}\gamma$, and QCD backgrounds requiring more than one particle to be misidentified. 

 In our cut-based analysis, we find an expected significance of $4.51 \pm 0.04~\sigma~ (4.70 \pm 0.04~\sigma)$ in the Regular (Improved) calorimeter resolution scenarios, and at the $68\%$ C.L., this sensitivity corresponds to a precision of  $46\%~(43\%)$ on the Higgs trilinear coupling, neglecting systematic uncertainties.   This is consistent with~\cite{Goncalves:2018yva} given the difference in backgrounds and analysis.  Further improvements could be made to this result if one was interested in improving the significance at $\lambda_3=1$, however, it is not clear how much further the precision of the coupling can be improved in this channel given the irreducible backgrounds.

Recent results from ATLAS and CMS with $36\,\mathrm{fb}^{-1}$ integrated luminosity indicate that while the $b\bar{b}\gamma\gamma$ channel is the cleanest, competitive limits can be set in the $b\bar{b}b\bar{b}$, $b\bar{b}\tau\tau$ and $b\bar{b}VV$ channels as well~\cite{Aaboud:2018knk, Aaboud:2018sfw, CMS-PAS-HIG-17-030}. Thus, assuming similar gains for these channels at $\sqrt{s} = 27\tev$, it's possible that a combined $\mathcal{O}(10-20\%)$ measurement of $\lambda_3$ may be possible at HE-LHC, making such a collider a powerful probe of electroweak symmetry breaking and BSM physics.


\subsection*{Acknowledgements}
We thank S. Jezequel and M. Wielers for help in understanding the ATLAS projections for High-Luminosity, and R. Contino for help understanding FCC projections.
We're also indebted to S. Dawson, G. Piacquadio and S. Prestel for useful conversations.
The work of S.H. and P.M.~was supported in part by NSF grant NSF-PHY-1620628.  
This work of P.M. was performed in part at the Aspen Center for Physics, which is supported by National Science Foundation grant PHY-1607611. P.M. would like to thank the Galileo Galilei Institute for Theoretical Physics for the hospitality and the INFN for partial support during the completion of this work, as well as support by a grant from the Simons Foundation (341344, LA).  Finally, P.M. would like to thank the Center for Theoretical Physics at Columbia University for hospitality during the completion of this work. 
This material is based upon work supported by the U.S. Department of Energy, Office of Science, Office of Workforce Development for Teachers and Scientists, Office of Science Graduate Student Research (SCGSR) program. The SCGSR program is administered by the Oak Ridge Institute for Science and Education (ORISE) for the DOE. ORISE is managed by ORAU under contract number DE-€SC0014664.


\clearpage
\appendix

\section{Validation with ATLAS Study}\label{a.validation}

In order to validate our performance assumptions for a detector at HE-LHC, and to keep consistent with the current projections for HL-LHC, we've made an effort where possible to align our definitions with the projections available from ATLAS~\cite{ATL-PHYS-PUB-2017-001}. This allows us to compare the results of our MC samples directly with the results presented in~\cite{ATL-PHYS-PUB-2017-001}, using the same cuts described in their paper. In this appendix we describe the results of this comparison, highlighting some of the discrepancies and improvements made in our study.

The full set of event selection criteria implemented in the ATLAS study are as follows:
\begin{itemize}
  \item $\geq 2$ isolated photons with $p_T > 30\gev$, $|\eta| < 1.37$ or $1.52 < |\eta| < 2.37$,
  \item $\geq 2$ jets identified as $b$-jets with leading/subleading $p_T > 40/30\gev$, $|\eta| < 2.4$,
  \item $< 6$ jets with $p_T > 30\gev$, $|\eta| < 2.5$,
  \item no isolated\footnote{The isolation criteria for leptons are not specified in~\cite{ATL-PHYS-PUB-2017-001}, so we assume them to be the same as the previous $hh\rightarrow b\bar{b}\gamma\gamma$ study~\cite{ATL-PHYS-PUB-2014-019}, where leptons were considered isolated if they had no $p_T > 10\gev$ jets in an annulus of $0.1 <  \Delta R < 0.4$. See, however, the discussion in the text regarding muons in $b$-jets.} leptons with $p_T > 25\gev$, $|\eta| < 2.5$
  \item $0.4 < \Delta R_{b\bar{b}}, \Delta R_{\gamma\gamma} < 2.0$, and $0.4 < \Delta R_{\gamma\text{jet}}$,
  \item $122 < m_{\gamma\gamma} < 128\gev$, $100 < m_{b\bar{b}} < 150\gev$,
  \item and $p_{T, \gamma\gamma}, p_{T, b\bar{b}} > 80\gev$.
\end{itemize}

\begin{table}[ht]
\centering
\begin{tabular}{|c|c c|}
\hline
Selection Requirement	& ATLAS Efficiency ($\%$)	& Our Efficiency ($\%$) \\
\hline
$\geq 2$ $b$-jet candidates											& 7.73	& 	7.61\\
$< 6$ jet candidates														& 7.46	& 	7.20\\
isolated lepton veto														& 6.96	& 	7.20\\
$0.4 < \Delta R_{b\bar{b}} < 2.0$, $\Delta R_{\gamma\gamma} < 2.0$ 	& 5.25	& 5.45\\
$122 < m_{\gamma\gamma} < 128 \gev$						  & 3.95	& 4.43\\
$100 < m_{b\bar{b}} < 150 \gev$									& 2.90	& 2.98\\
$h$ candidates $p_T > 80 \gev$									& 2.89	& 2.97\\
\hline
\end{tabular}
\caption{Cut flow for the signal process $hh\rightarrow b\bar{b}\gamma\gamma$ using the ATLAS event selection criteria. Shown are the results taken from~\cite{ATL-PHYS-PUB-2017-001} and the results obtained with the MC samples generated as described in \ssref{simulations} with the detector parameterization described in \ssref{detector}, using the HL-LHC benchmarks. Rows where the selection efficiency is not directly comparable between this work and~\cite{ATL-PHYS-PUB-2017-001} due to differences in the reweighting procedure are omitted.}
\label{t.atlas_cutflow}
\end{table}

The MC samples produced in our study (described in \ssref{simulations}) were subjected to the cuts summarized above. The expected number of events from each signal and background channel, at $3\,\mathrm{ab}^{-1}$ are listed in \tref{validation}, alongside the results of~\cite{ATL-PHYS-PUB-2017-001}, combining the \doubleq{barrel-barrel} and \doubleq{other} categorizations therein.

\begingroup
\renewcommand*{\arraystretch}{1.1}
\begin{table}[ht]
\centering
\begin{tabular}{|c|
  rcl
  rcl|}
\hline
\multicolumn{1}{|c|}{\multirow{2}{*}{Process}}  & \multicolumn{6}{c|}{Expected Events ($3\,\mathrm{ab}^{-1}$)} \\ \cline{2-7}
\multicolumn{1}{|c|}{}			& \multicolumn{3}{c|}{ATLAS~\cite{ATL-PHYS-PUB-2017-001}} & \multicolumn{3}{c|}{This Work}\\ \hline
$h(b\bar{b})h(\gamma\gamma)$	& $9.54$   & $\pm $ & $0.03$	& $9.91$  & $\pm$ & $0.04$       \\ \hline
$t\bar{t}h(\gamma\gamma)$	& $7.87$	& $\pm $ & $0.2$		& $10.1$  & $\pm$ & $0.1$    \\
$Zh(\gamma\gamma)$		& $4.98$	& $\pm $ & $0.1$		& $4.7$  & $\pm$ & $0.1$     \\
$b\bar{b}h(\gamma\gamma)$	& $0.15$	& $\pm $ & $0.01$		& $0.12$  & $\pm$ & $0.01$     \\
$ggF(\gamma\gamma)$		& $2.74$	& $\pm $ & $0.35$		& $14.3$  & $\pm$ & $0.6$     \\
$b\bar{b}\gamma\gamma$	& $21.80$	& $\pm $ & $0.6$		& $32.8$ & $\pm$ & $1.7$    \\
$c\bar{c}\gamma\gamma$	& $8.47$	& $\pm $ & $0.5$		& $9.1$  & $\pm$ & $0.3$     \\
$jj\gamma\gamma$			& $4.04$	& $\pm $ & $0.6$		& $3.9$  & $\pm$ & $0.2$     \\
$b\bar{b}j\gamma$			& $22.60$	& $\pm $ & $1.1$		& $17.0$ & $\pm$ & $1.9$    \\
$c\bar{c}j\gamma$			& $3.20$	& $\pm $ & $0.8$		& $3.4$  & $\pm$ & $0.4$     \\
$b\bar{b}jj$				& $5.35$	& $\pm $ & $0.8$		& $11.6$ & $\pm$ & $0.5$   \\
$Z(b\bar{b})\gamma\gamma$	& $2.06$	& $\pm $ & $0.1$		& $1.7$  & $\pm$ & $0.1$      \\
$t\bar{t}^{*}$				& $2.40$	& $\pm $ & $0.4$		& $2.5$  & $\pm$ & $0.5$     \\
$t\bar{t}\gamma$			& $5.16$	& $\pm $ & $0.5$		& $7.3$  & $\pm$ & $1.2$     \\ \hline
Total Background			& $90.82$	& $\pm $ & $2.0$		& $118.4$ & $\pm$ & $3.0$   \\ \hline
Significance ($S/\sqrt{B}$)  	& $1.00$	& $\pm $ & $0.02$		& $0.91$  & $\pm$ & $0.01$	\\ \hline
\end{tabular}
\caption{Comparison of the expected number of events from each signal/background process at $14\, {\rm TeV}$ with $3\,\mathrm{ab}^{-1}$ integrated luminosity from the ATLAS HL-LHC projections in~\cite{ATL-PHYS-PUB-2017-001} and this work, after imposing the selection criteria summarized in \aref{validation}. For the ATLAS expectations, we've combined the events categorized as \doubleq{barrel-barrel} and \doubleq{other}, which is consistent with our criteria. Also shown are the total background and significance (computed as $S/\sqrt{B}$) from each study.}
\label{t.validation}
\end{table}
\endgroup

In \tref{atlas_cutflow}, we compare the selection efficiency for the signal process ($hh\rightarrow b\bar{b}\gamma\gamma$) after each successive selection criterion from the ATLAS study is applied. There are two small differences, in the isolated lepton veto efficiency, and the efficiency of the $m_{b\bar{b}}$ cut.
After correspondence with the authors of~\cite{ATL-PHYS-PUB-2017-001}, it appears the discrepancy in the isolated lepton veto arises from muons in $b$-jets being erroneously labelled as isolated in the ATLAS study. Including the events which were discarded based on this isolation cut appears to resolve the discrepancy in the signal efficiency, however, it's unclear to what extent the additional isolated leptons suppressed the expected background numbers, making a precise comparison to our results difficult.

The difference in the $m_{b\bar{b}}$ cut efficiency can be attributed to differences in corrections made to $b$-jet four momenta in our study compared with the ATLAS study. The ATLAS work includes corrections such as the {\em muon-in-jet} and {\em PtReco} corrections, which increased the resolution of the $h\rightarrow b\bar{b}$ mass peak in previous ATLAS studies~\cite{Aaboud:2017xsd}. In contrast, we've approximated these corrections with only an additional energy scaling for $b$-jets, which corrects the $m_{b\bar{b}}$ peak location, but does not increase the resolution.

There are several differences between our setup and that used in the ATLAS study that should be noted when comparing the results.
The $ggF(\gamma\gamma)$ and $t\bar{t}$ backgrounds in the ATLAS study were generated using \PowhegBox, interfaced to \PythiaSix. In our study, these were instead produced in \MadGraphFull interfaced to \Pythia, with up to two extra partons (including $b$-jets) at the matrix element level. Additionally, the ATLAS study required the $t\bar{t}$ and $t\bar{t}\gamma$ samples to have at least one lepton in the final state at truth level, while we made no such requirement in our samples. The $t\bar{t}$ and $t\bar{t}\gamma$ samples were normalized using the same method in both studies. Finally, while the $b\bar{b}h(\gamma\gamma)$ background is included as part of the inclusive $ggF(\gamma\gamma)$ sample in our analysis, we also simulate it separately using the same methods used for $t\bar{t}h$ and $Zh$ to facilitate comparison with the ATLAS results.

The most significant differences between the two columns in \tref{validation} are in the $t\bar{t}h(\gamma\gamma)$, $ggF(\gamma\gamma)$, and $t\bar{t}\gamma$ backgrounds. As discussed above, the $t\bar{t}\gamma$ backgrounds in our study were not required to have a lepton, which likely explains the discrepancy. The $ggF$ backgrounds were also simulated differently in the two studies, which may explain in part the difference in the two projections. Finally, the $t\bar{t}h$ background in our sample is slightly higher, which is likely due to the presence of jets containing muons that were errantly vetoed in the ATLAS study. This is in fact a general trend, as nearly all of our signal and background estimates are somewhat higher on average than those in the ATLAS study, but agree with the estimates of~\cite{Chang:2018uwu}, which used the same cuts.

With the discussed above, the results in \tsref{atlas_cutflow}{validation} indicate that our \Delphes\ setup accurately describes a detector performance similar to the expectations at HL-LHC. We take this as confirmation that our setup also gives a reasonable projection to the $27\tev$ HE-LHC, with the HL-LHC performance as a useful benchmark

\clearpage

\bibliographystyle{JHEP}
\bibliography{triple}

\end{document}